\RequirePackage{amsthm}
\documentclass[sn-mathphys-num]{sn-jnl}

\usepackage{graphicx}%
\usepackage{multirow}%
\usepackage{amsmath,amssymb,amsfonts}%
\usepackage{amsthm}%
\usepackage{mathrsfs}%
\usepackage[title]{appendix}%
\usepackage{hyperref}
\usepackage{bookmark}
\usepackage{xcolor}%
\usepackage{textcomp}%
\usepackage{manyfoot}%
\usepackage{booktabs}%
\usepackage{algorithm}%
\usepackage{algorithmicx}%
\usepackage{algpseudocode}%
\usepackage{listings}%
\usepackage{amssymb}
\usepackage{amsmath}
\usepackage{amsfonts}
\usepackage{float}
\usepackage[caption=false]{subfig}
\usepackage[nameinlink]{cleveref}
\renewcommand\thesection{\arabic{section}}
\renewcommand\thesubsection{\arabic{section}.\arabic{subsection}}
\UseRawInputEncoding

\theoremstyle{thmstyleone}%

\theoremstyle{thmstyletwo}%

\theoremstyle{thmstylethree}%

\begin{document}

\title{Validity of energy conditions of matter in traversable wormholes under the $f(Q)$modified gravity theory}


\author*[1]{\sur{Jianbo Lu}}\email{lvjianbo819@163.com}

\author[1]{Shining Yang}

\author[1]{Yan Liu}

\author[1]{Yuying Zhang}

\author[1]{Yu Liu}

\affil*[1]{\orgdiv{Department of Physics}, \orgname{Liaoning Normal University}, \city{Dalian}, \postcode{116029},  \country{P. R. China}}


\abstract{ In the framework of the theory of general relativity, in order to obtain stable traversable wormholes, matter needs to violate the null energy condition. It is well known that the violation of the energy condition (EC) of matter leads to various physical problems. To address this issue, researchers have turned their attention to exploring modified theories of gravity, aiming to avoid the violation of ECs by introducing geometric terms. In this paper, within the framework of the $f(Q)$ modified gravitational theory, we investigate the effectiveness of ECs for matter in traversable wormholes. We examine the compliance of four types of energy conditions (weak energy condition, null energy condition, dominant energy condition, and strong energy condition) in the model by selecting a power-law model for $f(Q)$ and considering different shape functions $b(r)$. Our study reveals that for traversable wormholes realized through the $f(Q)$ modified gravity theory using the power-law model $f(Q)=a(-Q)^n$, all four types of ECs for matter can be satisfied. There is no need to introduce exotic matter (violating the null energy condition) or special matter (violating other energy conditions) artificially in the physics of wormholes.}




\maketitle

\section*{Introduction}\label{sec1}

Since the proposal of Einstein's theory of general relativity (GR), profound impacts have been made on the development of physics. Particularly, the confirmation of the existence of gravitational waves, as predicted in GR \cite{abbott2016observation}, has propelled the study of this theory to new heights. However, researchers face numerous difficulties and challenges when applying the gravitational theory of GR to solve related problems in cosmology and astrophysics, such as problems of thelate accelerated expansion of universe, the spacetime singularity, and the dark matter. Physicists attempt to solve the related problems existing in GR by modifying the gravitational theory \cite{nojiri2017modified,bull2016beyond,joyce2016dark}. As potential candidates for describing gravitational interaction, attempts have been made to propose feasible theories and establish relevant models from different perspectives, for example, $f(R)$ theory \cite{de2010f}, $f(T)$ theory \cite{bengochea2009dark}, and symmetric teleparallel gravity theory (STG) \cite{adak2006symmetric}. Since Nester and Yo first proposed STG theory \cite{1999Symmetric} in 1999, this theory has gradually gained attention. Extensive research has been carried out under this theory, such as the explorations of cosmology \cite{solanki2021cosmic,jimenez2020cosmology,khyllep2021cosmological}, gravitational wave physics \cite{soudi2019polarization}, and the geometric properties of wormholes \cite{banerjee2021wormhole}. In the symmetric teleparallel gravity theory, a connection with non-metricity is used to describe gravity, thereby neglecting the influence of curvature and torsion on spacetime, which geometrically implies that vectors do remain parallel at long distances on a manifold \cite{1999Symmetric}. It is well known that GR cannot distinguish between gravitation and inertial effects, but STG theory, as an equivalent description of GR, can distinguish these two forces through the frame fields \cite{koivisto2018integrable}. Unlike the fourth-order differential form of the field equations in $f(R)$ theory, $f(Q)$ modified gravity theory, as an extension of STG theory \cite{jimenez2018teleparallel}, owns field equations in the form of a more easily solvable second-order differential equation.

 A wormhole is a hypothetical structure of spacetime, functioning like a tunnel or bridge, connecting two asymptotically flat regions in a universe or two asymptotically flat universes \cite{bhattacharya2017f}. The earliest research on wormholes can be traced back to the Einstein-Rosen Bridge in the GR  \cite{einstein1935particle}. Additionally, wormhole solutions can be provided by transforming Kerr-Newman black hole solutions, but such solutions exhibit instability problems \cite{tian2015traversable,visser1995lorentzian}. In 1988, Morris and Thorne \cite{morris1988wormholes1} constructed the Morris-Thorne (MT) metric by introducing redshift function and shape function, exploring the physics of traversable wormholes. Since then, the study of the geometry of traversable wormholes officially began. In general relativity, one of the necessary conditions for forming a wormhole is that the matter needs to violate the null energy condition \cite{visser1995lorentzian}, which requires the introduction of exotic matter (i.e., matter violating the null energy condition) into wormhole physics. It is widely recognized that the violation of the energy conditions (ECs) of matter can induce and lead to some inconsistencies in physics, e.g., violation of the null energy condition (NEC) leads to stability problems in classical and quantum theory \cite{buniy2006null}; violation of the weak energy condition (WEC) leads to problem of negative energy density of fluid \cite{Lobo2002WeakEC}; violation of the dominant energy condition (DEC) implies that there is a superluminal problem in matter \cite{schoen1981proof}; and violation of the strong energy condition implies that the attraction property of gravity is destroyed, and Hawking Penrose singularity theories will fail \cite{santos2006energy,moraes2017analytical,Hawking2023TheLS}, etc. Consequently, physicists have conducted extensive research, exploring the realization of stable traversable wormholes under the condition that matter does not violate the energy condition. Typically, the realization of traversable wormhole physics is examined within the framework of modified gravity theory \cite{mehdizadeh2015einstein,mazharimousavi2016wormhole,bronnikov2003possible} (through the effect of geometric terms), in order to maximize the validity of the energy condition of matter. In this paper, we study wormhole physics within the framework of $f(Q)$ modified gravity theory.

 This paper is divided into five sections.  The first chapter is the introduction. In the second chapter, we briefly introduce the basic equations of $f(Q)$ modified gravity theory. The third chapter outlines the energy conditions of matter and the basic conditions for forming a traversable wormhole. In the fourth chapter, we explore the realization of traversable wormholes that satisfy the energy conditions of normal matter within the framework of $f(Q)$ modified gravity. Section V is the conclusion.

\section*{The basic equations of $f(Q)$ symmetric teleparallel modified theory of gravity}\label{sec2}

In the $f(Q)$ modified gravity theory, gravitational interactions are described by the non-metricity term $Q$, and its action can be expressed as follows \cite{jimenez2018teleparallel}:
\begin{equation}
 S=\int \frac{1}{2} f(Q) \sqrt{-g} d^4 x+\int \mathcal{L}_m \sqrt{-g} d^4 x,\label{2.1}
\end{equation}
 where $f(Q)$ represents an arbitrary function of $Q$, $g$ is the determinant of the metric $g_{\mu v}$, and $\mathcal{L}_m$ is the Lagrangian density of matter.
 The non-metricity tensor can be written as:
\begin{equation}
 Q_{\lambda \mu \nu}=\nabla_{\lambda} g_{\mu \nu}.\label{2.2}
\end{equation}
 The non-metricity tensor has two independent traces, which can be represented as:
\begin{equation}
 Q_\alpha=Q_{\alpha~\mu}^{~\mu}, \tilde{Q}_\alpha=Q^\mu_{~\alpha \mu}.\label{2.3}
\end{equation}
 The non-metricity scalar is defined as:
 \begin{equation}
  Q=-Q_{\alpha \mu v} P^{\alpha \mu v}.\label{2.4}
  \end{equation}
 By utilizing the non-metricity tensor, the superpotential can be written as:
 \begin{equation}
  P_{~\mu \nu}^\alpha=\frac{1}{4}\left[-Q_{~\mu \nu}^\alpha+2 Q_{(\mu\nu)}^\alpha+Q^\alpha g_{\mu \nu}-\widetilde{Q}^\alpha g_{\mu \nu}-\delta_{(\mu}^\alpha Q_{v)}\right].\label{2.5}
  \end{equation}
 The energy-momentum tensor is represented as:
 \begin{equation}
  T_{\mu \nu}=-\frac{2}{\sqrt{-g}} \frac{\delta\left(\sqrt{-g} \mathcal{L}_m\right)}{\delta g^{\mu \nu}}.\label{2.6}
 \end{equation}
 Taking the variation with respect to the metric tensor $g_{\mu v}$ in equation (\ref{2.1}) yields the field equation
 \begin{equation}
 \begin{gathered}
 \frac{2}{\sqrt{-g}} \nabla_\alpha\left(\sqrt{-g} f_Q P_{~~\mu \nu}^{\alpha}\right)+\frac{1}{2} g_{\mu \nu} f +f_Q\left(P_{\mu\alpha\beta} Q_{\nu}^{~\alpha\beta}-2 Q_{\alpha\beta\mu} P^{\alpha\beta}_{~~~\nu}\right)=-T_{\mu \nu},\label{2.7}
 \end{gathered}
 \end{equation}
 where $f_Q=\frac{d f}{d Q}$. Furthermore, the variation with respect to the connection yields the following relationship
 \begin{equation}
  \nabla_\mu \nabla_v\left(\sqrt{-g} f_Q P_{~~~\alpha}^{\mu\nu}\right)=0.\label{2.8}
  \end{equation}

\section*{Traversable wormholes and energy conditions}\label{sec3}

The spacetime of wormholes under static spherical symmetry can be described by the MT line element, with its form as follows \cite{morris1988wormholes1}:
 \begin{equation}
 d s^2=-e^{2 \phi(r)} d t^2+\left(1-\frac{b(r)}{r}\right)^{-1} d r^2+r^2 d \theta^2+r^2 \sin ^2 \theta d \varphi^2,\label{3.1}
 \end{equation}
 where $\phi(r)$ and $b(r)$ are functions of the radial coordinate $r$, and the range of $r$ is $r_0 \rightarrow \infty$, with $r_0$ representing the position of the wormhole throat. $\phi(r)$ is referred to as the redshift function. $b(r)$ represents the shape function, the form of which determines the geometric shape of the wormhole. To ensure that the line element (\ref{3.1}) can describe a stable wormhole, $b(r)$ needs to meet the following conditions:

(1) Throat condition: $b(r_0 )=r_0$, and  $b(r)$ should be less than $r$ for $r>r_0$;

(2) Flaring-out condition: $b^{\prime}\left(r_0\right)<1$ and $\frac{b(r)-r b^{\prime}(r)}{b^2(r)}>0$;

(3) Asymptotically flat condition: when $r\rightarrow \infty$, $\frac{b(r)}{r} \rightarrow 0$.

 Furthermore, to avoid the event horizon, it is necessary to require that the redshift function $\phi(r)$ is finite everywhere (i.e., $e^{2 \phi} \neq 0$) for a traversable wormhole. The distance measure $l(r)$ is defined as follows \cite{sokoliuk2022traversable}:
\begin{equation}
 l(r)= \pm \int_{r_0}^r \frac{d r}{\sqrt{1-\frac{b(r)}{r}}}.\label{3.2}
\end{equation}
 Another important condition for the formation of a traversable wormhole is that $l(r)$ must be finite on the radial coordinate and should be a decreasing function. Furthermore, $l$ should be larger than or equal to the radial coordinate distance, i.e., $\mid l(r) \mid \geq r-r_0$. The positive and negative signs of $l$ respectively represent the upper and lower parts of the wormhole, which are connected through the wormhole throat.

 Considering that the matter of a traversable wormhole is described by an anisotropic energy-momentum tensor, its specific form is as follows:
 \begin{equation}
 T_\mu^v=\left(\rho+P_t\right) u_\mu u^v-P_t \delta_\mu^v+\left(P_r-P_t\right) v_\mu v^v.\label{3.3}
 \end{equation}
 where, $u_\mu$ is the four-velocity vector, $v_\mu$ is the radial spacelike vector. $\rho$ denotes the energy density, $P_r$ and $P_t$ respectively represent the pressure along the direction of $u_\mu$ (radial pressure) and the pressure in the direction orthogonal to $v_\mu$ (tangential pressure).

 In the exploration of modified gravitational theories, study on energy conditions is a hot topic that people often discuss. It is hoped to solve the problem of violated ECs of matter through the modification of gravitational theories. For example, in GR, to solve problems such as the early inflation \cite{baldi2005inflation} and late accelerating expansion \cite{santos2007energy} of the universe, it is necessary to introduce special matter-dark energy, which violates the strong energy condition. It is well known that, the violation of relevant energy conditions usually induce some problems in physics, e.g., which could challenge the stability in classical and quantum theory \cite{buniy2006null}, bring about the ill-defined negative energy density of fluid \cite{Lobo2002WeakEC}, indicate the appearance of superluminal matter \cite{schoen1981proof}, destroy the attraction property of gravity and Hawking Penrose singularity theories \cite{santos2006energy,moraes2017analytical,Hawking2023TheLS}, etc.  However, it has been found in research that in GR, matter must violate NEC to form a traversable wormhole. In fact, in this theory, the WEC, DEC, and SEC of matter all need to be violated \cite{visser1995lorentzian,lobo2008general}. Therefore, exploring the energy conditions of matter in traversable wormhole under the framework of modified gravitational theories is extremely important. In this aspect, lots of works have been carried out, for example:

 (1) In $f(R)$ modified gravitational theory, Ref.\cite{julianto2022traversable} assumes that the equation of state of matter satisfies the relationship $P_r=\omega \rho$. Basing on the derived $f(R)$ model and considering different shape functions ($b(r)=r_0\left(\frac{r}{r_0}\right)^{\frac{1}{2}}$,or $b(r)=\frac{r_0 \log (1+r)}{\log \left(1+r_0\right)}$), it is found that the matter in traversable wormhole satisfies the tangential NEC and violates the radial NEC.

 (2) In the generalized Brans-Dicke model \cite{lu2023traversable}, under the consideration of two different equations of state ($P_r=\omega \rho$, or $P_t=\omega \rho$) and a specific shape function ($b(r)=r_0\left(\frac{r}{r_0}\right)^A$), it can be observed that the traversable wormhole matter satisfies NEC, WEC, and DEC, but does not satisfy SEC.

 (3) In $f(Q)$ theory of modified gravity \cite{hassan2021traversable}, the authors use a linear $f(Q)$ model ($f(Q)=\alpha Q$) and conditions (e.g., the equation of state $P_t=mP_r$ and $P_r=\omega \rho$) to solve the field equations. The results show that traversable wormhole matter needs to violate NEC.

 In these modified gravitational theories and other studies mentioned above, it is easy to see that the realization of stable traversable wormholes still needs to introduce exotic matter or some special matter that violates relevant energy conditions to maintain. Obviously, this result is unsatisfactory. Therefore, exploring the geometry of traversable wormholes that does not depend on exotic or special matter is a hot issue in current research and is of important theoretical significance.

 ECs are derived from the Raychaudhuri equation. This equation is exclusively geometric which is not deal to any theory of gravity \cite{sharif2013energy}, and then it has been utilized to explore the properties of matter in many theories, such as, $f(R)$ theory \cite{bertolami2009energy,capozziello2018role,santos2017strong}, $f(T)$ theory \cite{zubair2015energy}, $f(R,\tilde{T})$ theory \cite{zubair2016static,chanda2021morris,zubair2022energy},  $f(Q)$ theory \cite{mandal2020energy,koussour2022cosmic,mustafa2021wormhole}, and $f(R,\phi)$ theory \cite{zubair2018static}, etc.  The Raychaudhuri equation describes the temporal evolution of expansion scalar ($\theta$) for the congruences of timelike ($u^\mu$) and null ($\eta^\mu$) geodesics as \cite{raychaudhuri1955relativistic}:
 \begin{equation}
 \frac{d \theta}{d \tau}-\omega_{\mu \nu} \omega^{\mu \nu}+\sigma_{\mu \nu} \sigma^{\mu \nu}+\frac{1}{3} \theta^2+R_{\mu \nu} u^\mu u^\nu=0 \label{3.4}
 \end{equation}
 \begin{equation}
 \frac{d \theta}{d \tau}-\omega_{\mu \nu} \omega^{\mu \nu}+\sigma_{\mu \nu} \sigma^{\mu \nu}+\frac{1}{2} \theta^2+R_{\mu \nu} \eta^\mu \eta^\nu=0.\label{3.5}
 \end{equation}
 Here $\sigma^{\mu v}$ and $\omega_{\mu v}$ are the shear and rotation related to the vector field, respectively. The shear satisfies relation: $\sigma^2=\sigma_{\mu \nu} \sigma^{\mu \nu} \geq 0$ and the expansion scalar has: $\theta^{2}\geq 0$. One can simplify to derive the resulting inequalities by neglecting the quadratic terms in equations (\ref{3.4}) and (\ref{3.5}). Concretely, it can be assumed that there are infinitesimal distortions in geodesics which is hypersurface orthogonal as well  \cite{zubair2015energy}, i.e., $\omega_{\mu\nu}=0$ (irrotational congruences), then the integration of simplified Raychaudhari equations results in $\theta=-\tau R_{\mu\nu}u^\mu u^{\nu}=-\tau R_{\mu\nu}\eta^\mu \eta^{\nu}$  for timelike and null geodesics, respectively. Furthermore, considering the attractive nature of gravity with utilizing $\frac{d\theta}{d\tau}<0$, one can get
 \begin{equation}
 R_{\mu \nu} u^\mu u^\nu \geq 0\label{3.6}
 \end{equation}
 \begin{equation}
  R_{\mu \nu} \eta^\mu \eta^\nu \geq 0.\label{3.7}
  \end{equation}

Considering the traversable wormhole matter is anisotropical distribution, the energy-condition relations can be derived as follows \cite{morris1988wormholes1,morris1988wormholes2}:

1. Null Energy Condition: $\rho+P_r \geq 0 ,\  \rho+P_t \geq 0$,

2. Weak Energy Condition: $\rho \geq 0 ,\ \rho+P_r \geq 0 ,\ \rho+P_t \geq 0$,

3. Dominant Energy Condition: $\rho \geq 0 ,\ \rho \pm P_r \geq 0 ,\ \rho \pm P_t \geq 0$,

4. Strong Energy Condition: $\rho+P_r \geq 0 ,\ \rho+P_t \geq 0 ,\ \rho+P_r+2 P_t \geq 0$.

We can discuss  the stability and consistency of theoretical models by studying the energy conditions of matter \cite{sharma2022traversable}.

\section*{Stable traversable wormholes under $f(Q)$ modified gravity theory}\label{sec4}

 In this section, we investigate the properties of traversable wormhole matter by examining the four energy conditions under the framework of $f(Q)$ gravity. Under the MT line element, the trace of the non-metric tensor $Q$ is written as \cite{mustafa2021wormhole}:
 \begin{equation}
 Q=-\frac{2}{r}\left(1-\frac{b(r)}{r}\right)\left(2 \phi^{\prime}(r)+\frac{1}{r}\right).\label{4.1}
 \end{equation}
 Substituting equations (\ref{3.1}) and (\ref{3.3}) into equation (\ref{2.7}) yields the following relations:
 \begin{equation}
 \left[\frac{1}{r}\left(-\frac{1}{r}+\frac{r b^{\prime}(r)+b(r)}{r^2}-2 \phi^{\prime}(r)\left(1-\frac{b(r)}{r}\right)\right)\right] f_Q-\frac{2}{r}\left(1-\frac{b(r)}{r}\right) \dot{f}_{\mathrm{Q}}-\frac{f}{2}=-\rho,\label{4.2}
 \end{equation}
 \begin{equation}
 \left[\frac{2}{r}\left(1-\frac{b(r)}{r}\right)\left(2 \phi^{\prime}(r)+\frac{1}{r}\right)-\frac{1}{r^2}\right] f_Q+\frac{f}{2}=-P_r,\label{4.3}
 \end{equation}
 \begin{equation}
 \begin{aligned}
  & {\left[\frac{1}{r}\left(\left(1-\frac{b(r)}{r}\right)\left(\frac{1}{r}+\phi^{\prime}(r)\left(3+r \phi^{\prime}(r)\right)+r \phi^{\prime \prime}(r)\right)-\frac{r b^{\prime}(r)-b(r)}{2 r^2}\left(1+r \phi^{\prime}(r)\right)\right)\right] f_Q+} \\
    & \frac{1}{r}\left(1-\frac{b(r)}{r}\right)\left(1+r \phi^{\prime}(r)\right) \dot{f}_{\mathrm{Q}}+\frac{f}{2}=-P_t,\label{4.4}
    \end{aligned}
    \end{equation}
 where $\rho$, $P_r$  and $P_t$ represent the energy density, radial pressure  and tangential pressure of matter in the wormhole, respectively.

 We consider that the redshift function $\phi(r)$ is finite and does not vanish at the wormhole's throat. For convenience in calculation, we take $\phi(r)$ as a constant. At this point, the form of equation (\ref{4.1}) becomes:
 \begin{equation}
  Q=\frac{2}{r^2}\left(\frac{b(r)}{r}-1\right).\label{4.5}
  \end{equation}
 After applying the condition $\phi^{\prime}(r)=0$ to equations (\ref{4.2})-(\ref{4.4}) and simplifying, we can obtain the specific forms of energy density, radial pressure, and tangential pressure as follows:
 \begin{equation}
 \rho=\left(\frac{1}{r^2}-\frac{r b^{\prime}(r)+b(r)}{r^3}\right) f_Q+\frac{2}{r}\left(1-\frac{b(r)}{r}\right) \dot{f}_{\mathrm{Q}}+\frac{f}{2},\label{4.6}
 \end{equation}
 \begin{equation}
 P_r=\left(\frac{2 b(r)}{r^3}-\frac{1}{r^2}\right) f_Q-\frac{f}{2},\label{4.7}
  \end{equation}
  \begin{equation}
   P_t=\left(\frac{b^{\prime}(r)}{2 r^2}+\frac{b(r)}{2 r^3}-\frac{1}{r^2}\right) f_Q-\left(\frac{1}{r}-\frac{b(r)}{r^2}\right) \dot{f}_{\mathrm{Q}}-\frac{f}{2}.\label{4.8}
    \end{equation}
 To continue with the calculation, we select a specific power-law model \cite{gadbail2023cosmology,koussour2023constant,sarmah2023anisotropic} of $f(Q)$: $f(Q)=a(Q)^n$ (where $a$ and $n$ are model parameters), and consider studying the properties of the constructed wormhole physical model under two different forms of shape functions. This model has been used by researchers to test its properties  and values in several fields. For example, Refs.\cite{gadbail2023cosmology,koussour2023constant} utilized it to investigate the interpretation  of the late-stage accelerated expansion of the universe, while Ref.\cite{sarmah2023anisotropic} discussed its related properties in the early universe. In this paper, we explore the physical properties of the traversable wormhole under this model. By substituting $f(Q)=a(Q)^n$ into equations (\ref{4.6})-(\ref{4.8}), we obtain the expressions for the energy density, radial pressure, and tangential pressure of the matter respectively \cite{sokoliuk2022traversable}:
\begin{equation}
\rho=a\left[\frac{n r b^{\prime}(r)+(n-1)(b(r)-r)}{r^3}+n(n-1) r\right]\left(\frac{2(r-b(r))}{r^3}\right)^{n-1},\label{4.9}
\end{equation}
\begin{equation}
 P_r=-\frac{a}{2}\left(n \frac{2 b(r)-r}{r-b(r)}+1\right)\left(\frac{2(r-b(r))}{r^3}\right)^n,\label{4.10}
 \end{equation}
 \begin{equation}
  P_t=-\frac{a}{2}\left[\frac{n r b^{\prime}(r)+(n-2) b(r)-2(n-1) r}{r^3}+n(n-1) r\right]\left(\frac{2(r-b(r))}{r^3}\right)^{n-1}.\label{4.11}
 \end{equation}

 \subsection*{\text{Shape function: $\mathrm{b}(\mathrm{r})=\mathrm{r}_0\left(\frac{\mathrm{r}}{\mathrm{r}_0}\right)^{\mathrm{m}}$}}

\begin{figure}[ht]
    \renewcommand{\thefigure}{1}
    \centering
    \subfloat{\includegraphics[width=0.5\hsize]{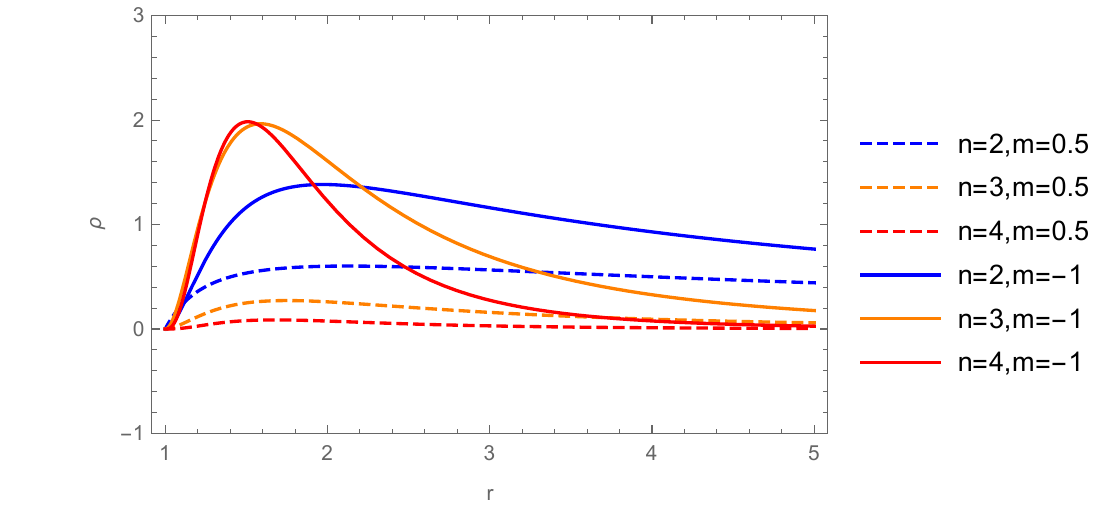}}
    \subfloat{\includegraphics[width=0.5\hsize]{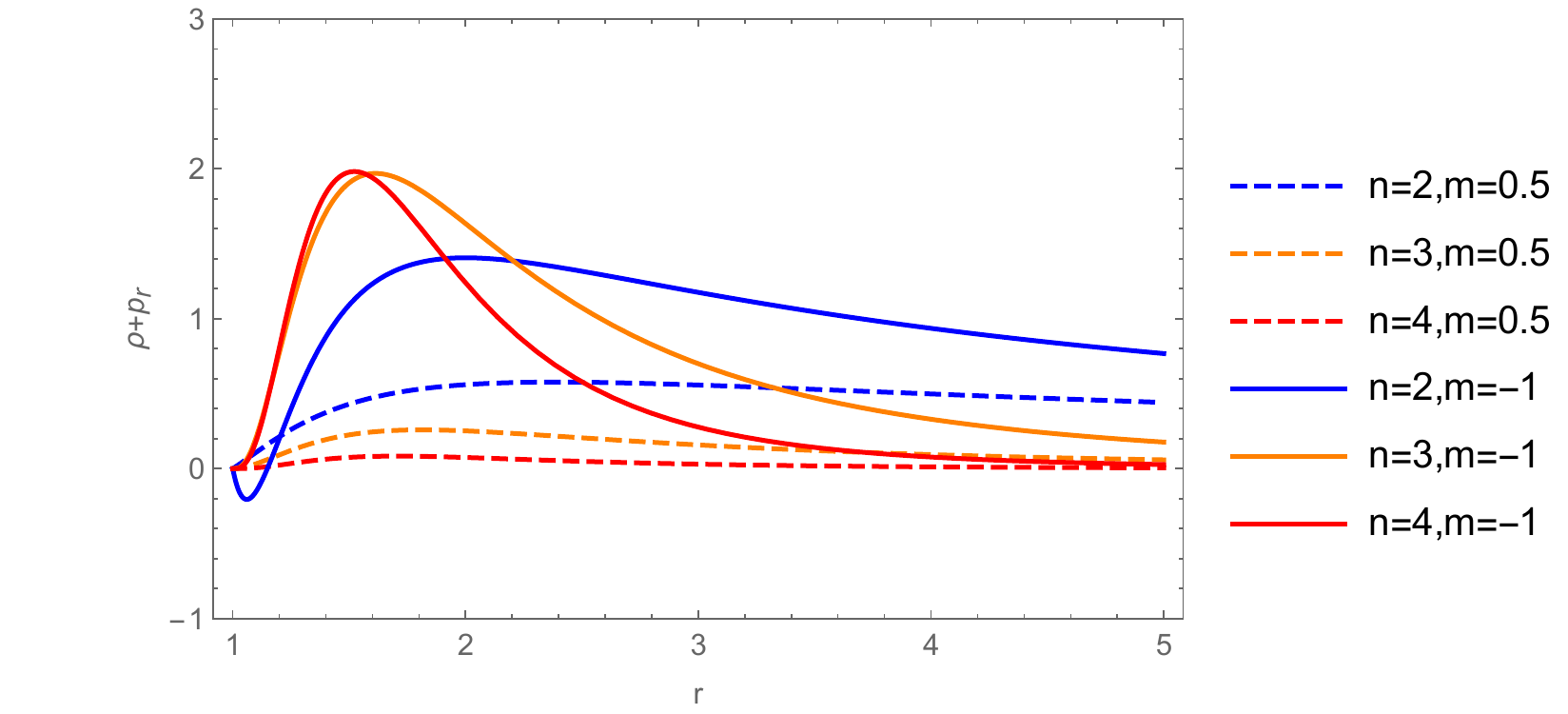}}
    \vspace{2mm}\\
    \subfloat{\includegraphics[width=0.5\hsize]{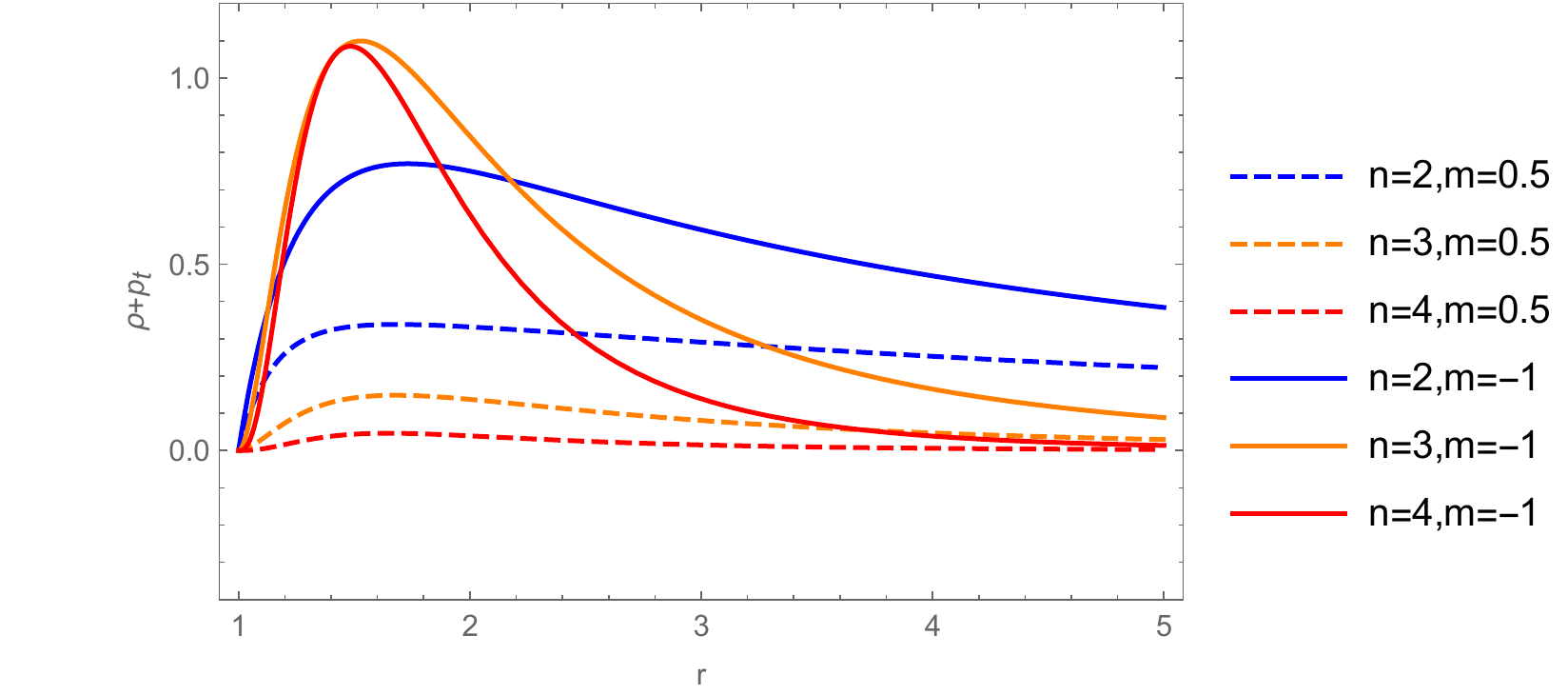}}
    %
    \subfloat{\includegraphics[width=0.5\hsize]{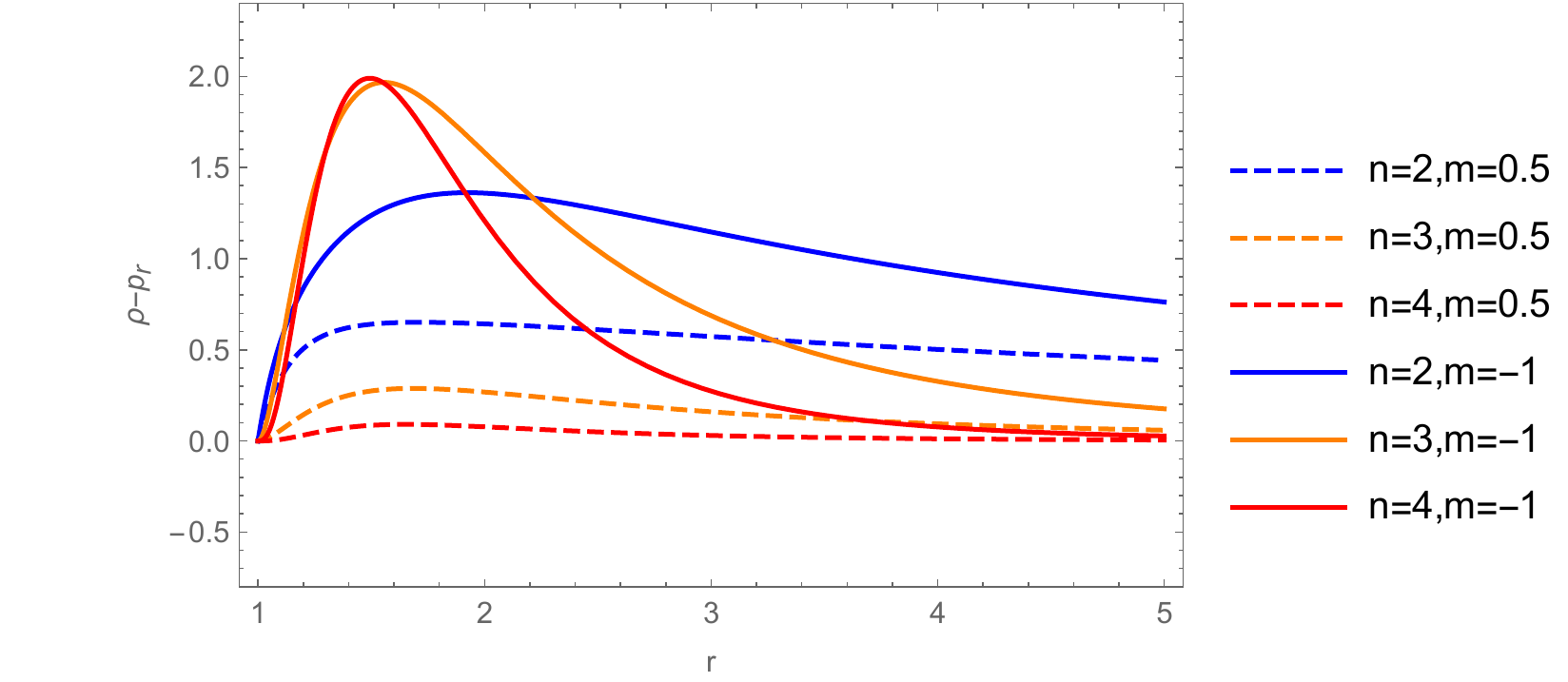}}
    \vspace{2mm}\\
    \subfloat{\includegraphics[width=0.5\hsize]{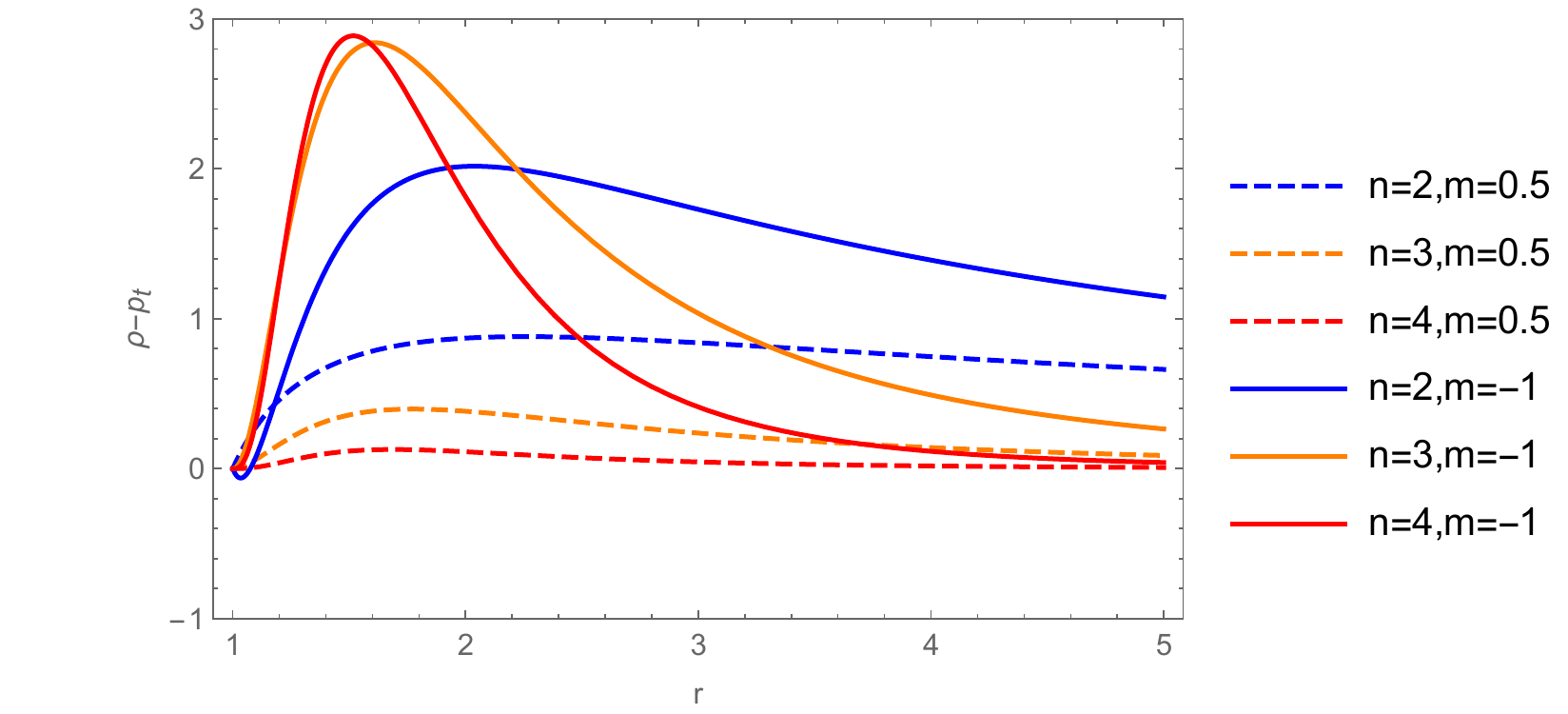}}
    \subfloat{\includegraphics[width=0.5\hsize]{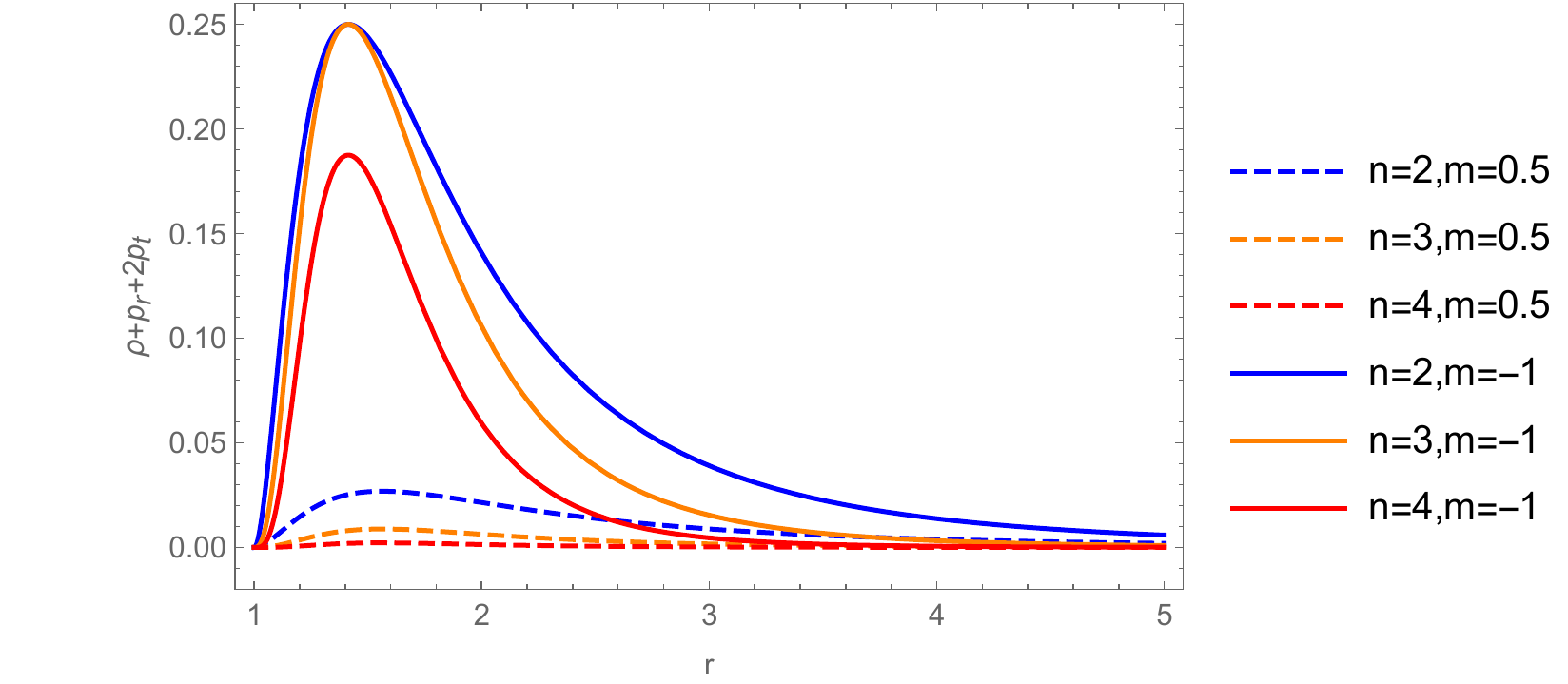}}
    \caption{Variation of the energy condition relations relative to the radial function $r$ in the power-law $f(Q)$  model, where the shape function is $b(r)=r_0\left(\frac{r}{r_0}\right)^{m}$.} \label{fig:4.1}
\end{figure}

 We consider a power-law model of shape function, i.e., $b(r)=r_0\left(\frac{r}{r_0}\right)^{m}$, where $r_0$ and $m$ are constant parameters. As a parameterized shape function, this model has been widely used in traversable wormholes for studying the modified gravitational theories \cite{julianto2022traversable,lu2023traversable,hassan2022embedding,cataldo2017traversable,luis2022non}. It is easy to find that under this shape function, the energy conditions of matter in the traversable wormhole can not meet all energy conditions in the relevant  modified gravities. By substituting shape function $b(r)=r_0\left(\frac{r}{r_0}\right)^{m}$ into equations (\ref{4.9})-(\ref{4.11}) of $f(Q)$ theory, the specific expressions for the four energy conditions can be written as follows:
 \begin{equation}
 \rho=a\left[\frac{(m n+n-1) r^{m-3}}{r_0^{m-1}}-\frac{n-1}{r^2}+n(n-1) r\right]\left(\frac{2}{r^2}-\frac{2 r^{m-3}}{r_0^{m-1}}\right)^{n-1},\label{4.12}
 \end{equation}
 \begin{equation}
 \rho+P_r=a n\left[(m-1) \frac{r^{m-3}}{r_0^{m-1}}+(n-1) r\right]\left(\frac{2}{r^2}-\frac{2 r^{m-3}}{r_0^{m-1}}\right)^{n-1},\label{4.13}
 \end{equation}
 \begin{equation}
 \rho+P_t=a \frac{n}{2}\left[\frac{(m+1) r^{m-3}}{r_0^{m-1}}+(n-1) r\right]\left(\frac{2}{r^2}-\frac{2 r^{m-3}}{r_0^{m-1}}\right)^{n-1},\label{4.14}
 \end{equation}
 \begin{equation}
 \rho-P_r=a\left[\frac{(m n+3 n-2) r^{m-3}}{r_0^{m-1}}+\frac{2(1-n)}{r^2}+n(n-1) r\right]\left(\frac{2}{r^2}-\frac{2 r^{m-3}}{r_0^{m-1}}\right)^{n-1},\label{4.15}
 \end{equation}
\begin{equation}
\rho-P_t=a\left[\frac{(3 m n+3 n-4) r^{m-3}}{2 r_0^{m-1}}+\frac{2(1-n)}{r^2}+\frac{3}{2} n(n-1) r\right]\left(\frac{2}{r^2}-\frac{2 r^{m-3}}{r_0^{m-1}}\right)^{n-1},\label{4.16}
\end{equation}
 \begin{equation}
 \rho+P_r+2 P_t=2^n a(n-1)\left(\frac{r-r^m r_0^{1-m}}{r^3}\right)^n.\label{4.17}
 \end{equation}
 To intuitively observe the variations of energy conditions relative to the radial coordinate, we perform numerical calculations on the above expressions and display them in graphical form. The parameter values are taken as follows. Without loss of generality, we choose the parameter (denoting the position of wormhole throat) $r_0=1$. In addition, from equations (\ref{4.12})-(\ref{4.17}) it is easy to see that the value of the constant parameter $a$ only affects the magnitude of the function values of the energy conditions (it appears as a proportion factor in the expressions, and does not affect the positivity or negativity of the function values). Thus, we can choose the value of parameter $a=1$. Furthermore, in light of the condition that as $r \rightarrow \infty, \frac{b(r)}{r} \rightarrow 0$, the model parameter $m$ must meet the restriction condition: $m<1$. In this paper, we consider two situations where $m$ is positive ($m=1/2$) and negative ($m=-1$) for study. Specifically, we plot the variation curves of the four types of energy conditions (NEC, WEC, DEC, SEC) relative to the radial coordinate $r$ for both cases of $m=1/2$ (represented by solid lines) and $m=-1$ (represented by dashed lines) in \Cref{fig:4.1} (the model parameter $n$ values are taken as $2, 3$ and  $4$, respectively).

\begin{figure}[ht]
    \renewcommand{\thefigure}{2}
    \centering
    \subfloat{\includegraphics[width=0.5\hsize]{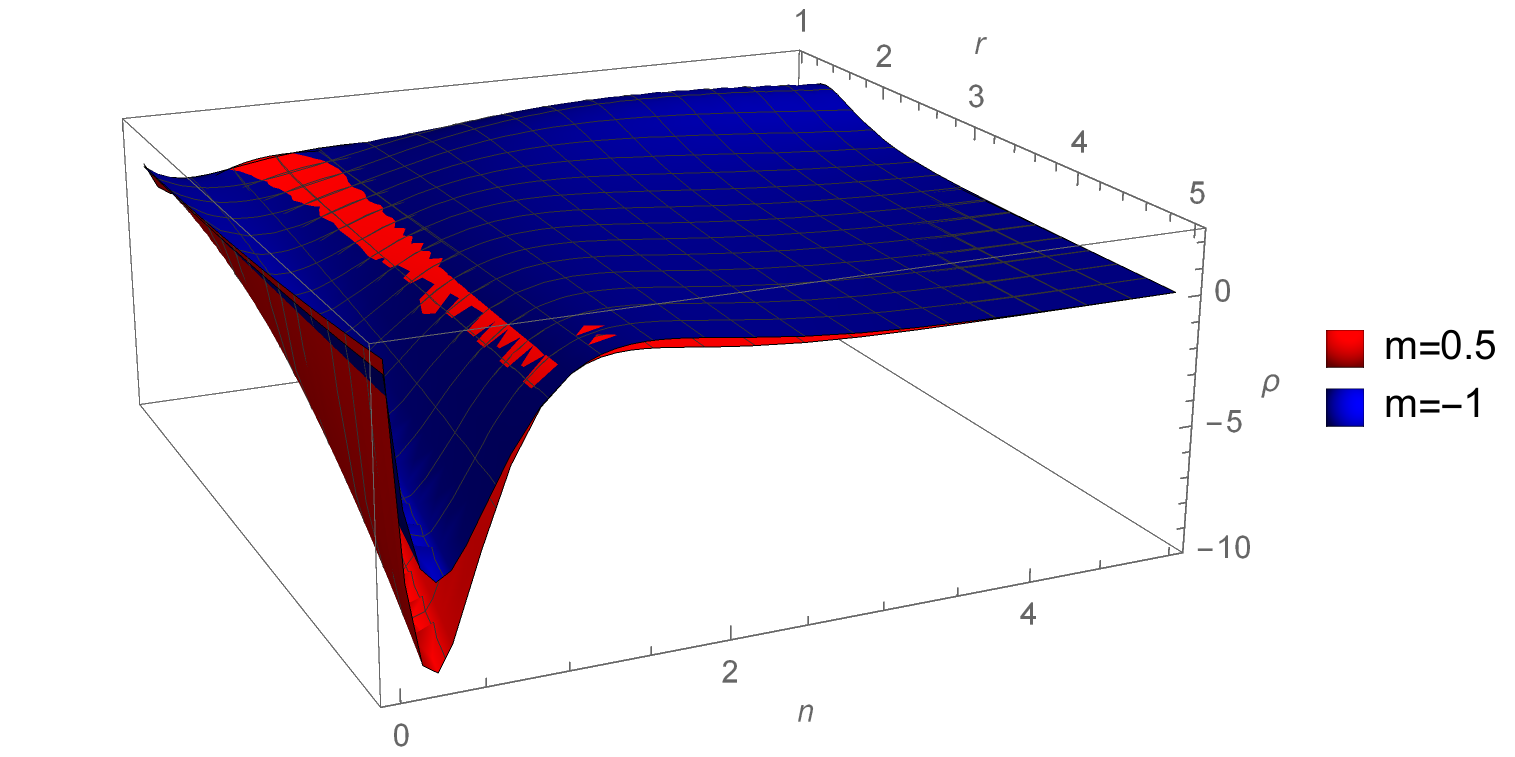}}
    \subfloat{\includegraphics[width=0.5\hsize]{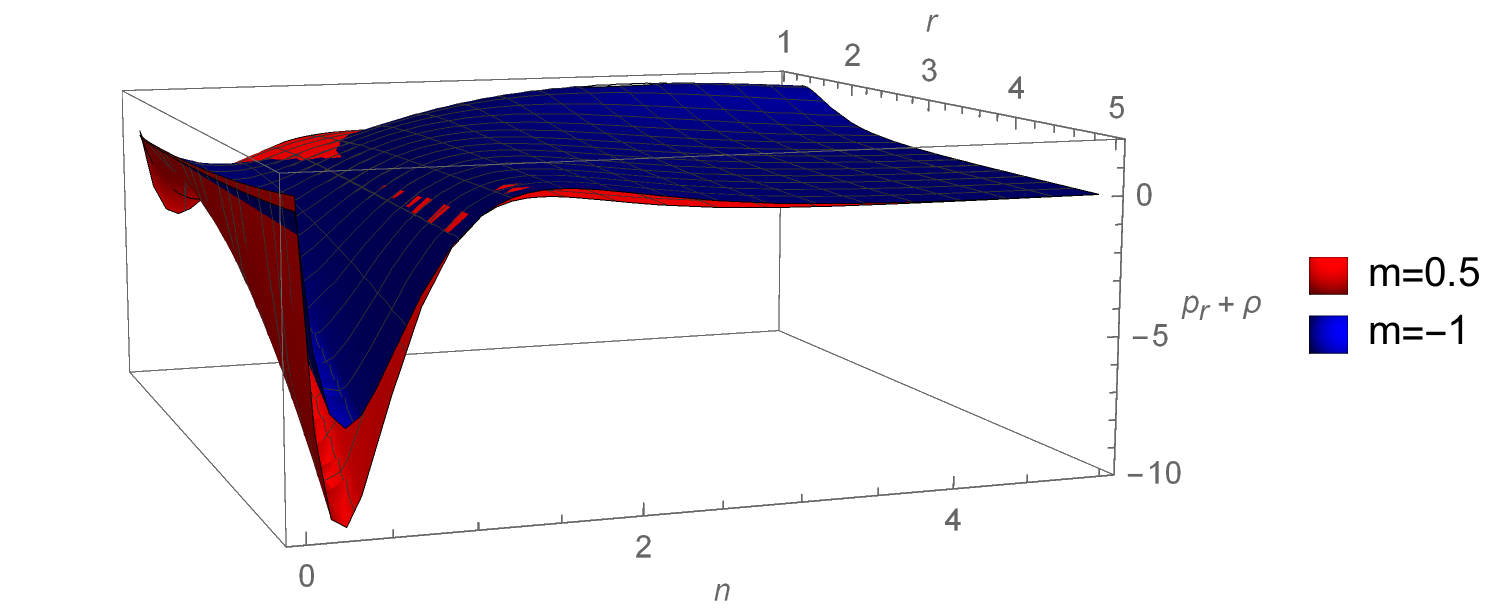}}
    \\
    \subfloat{\includegraphics[width=0.5\hsize]{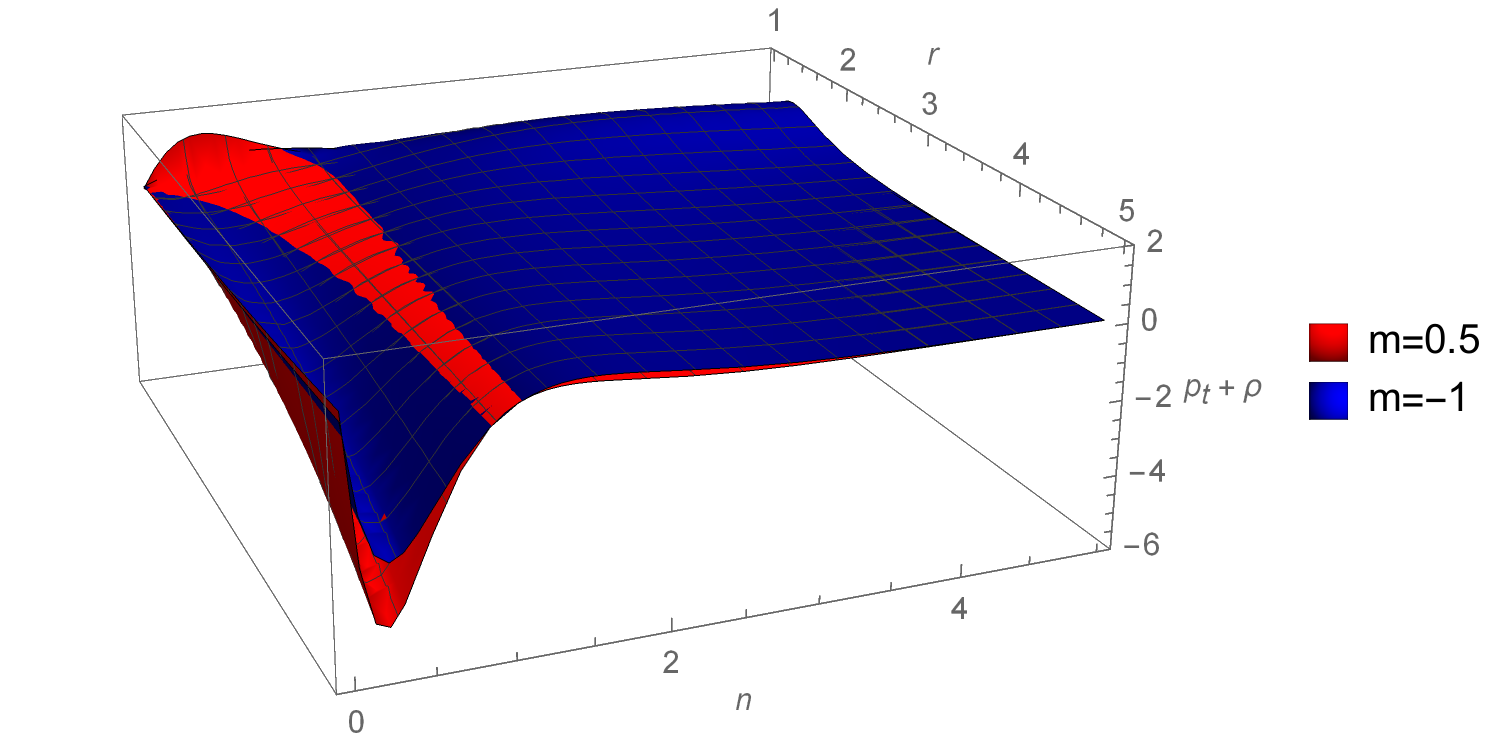}} 
    \subfloat{\includegraphics[width=0.5\hsize]{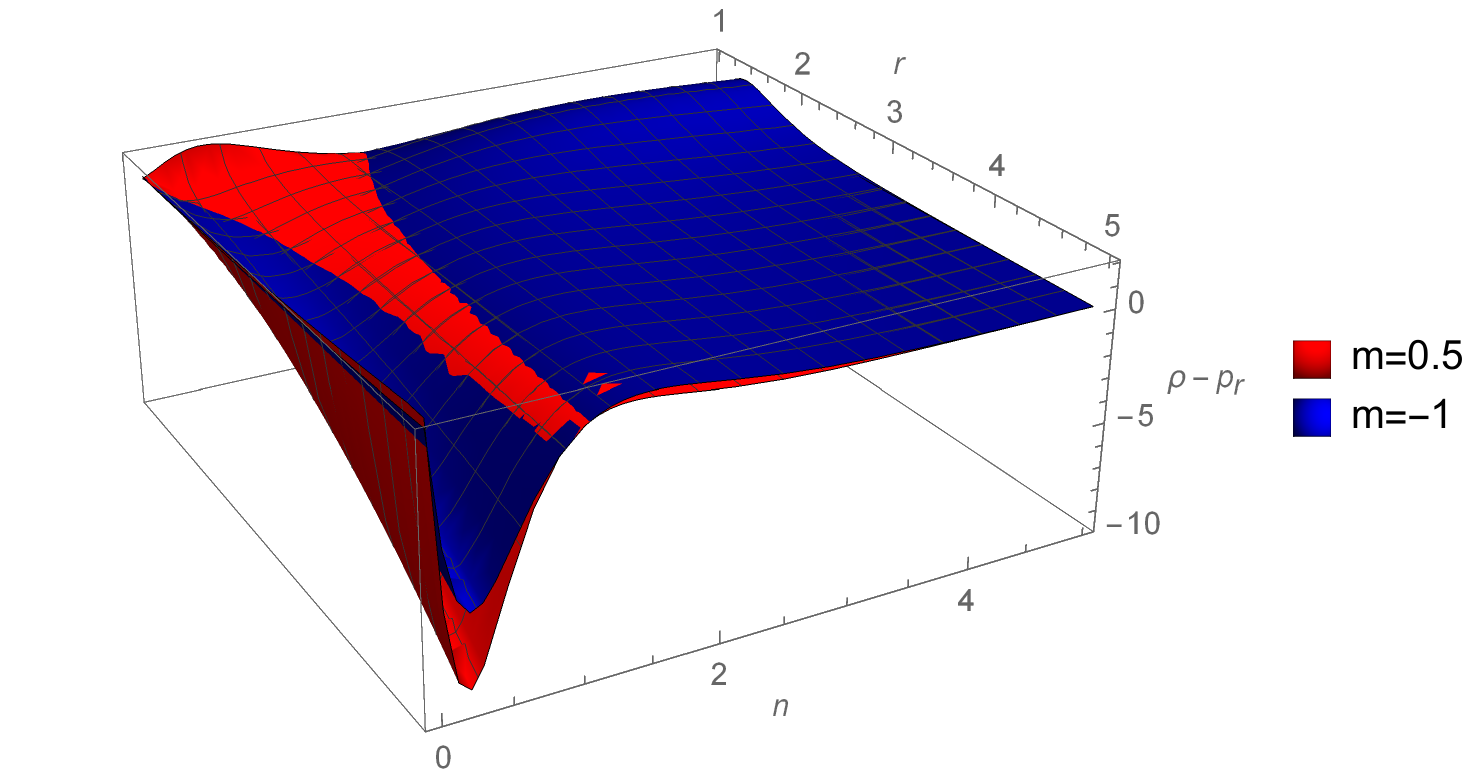}}
    \\
    \subfloat{\includegraphics[width=0.5\hsize]{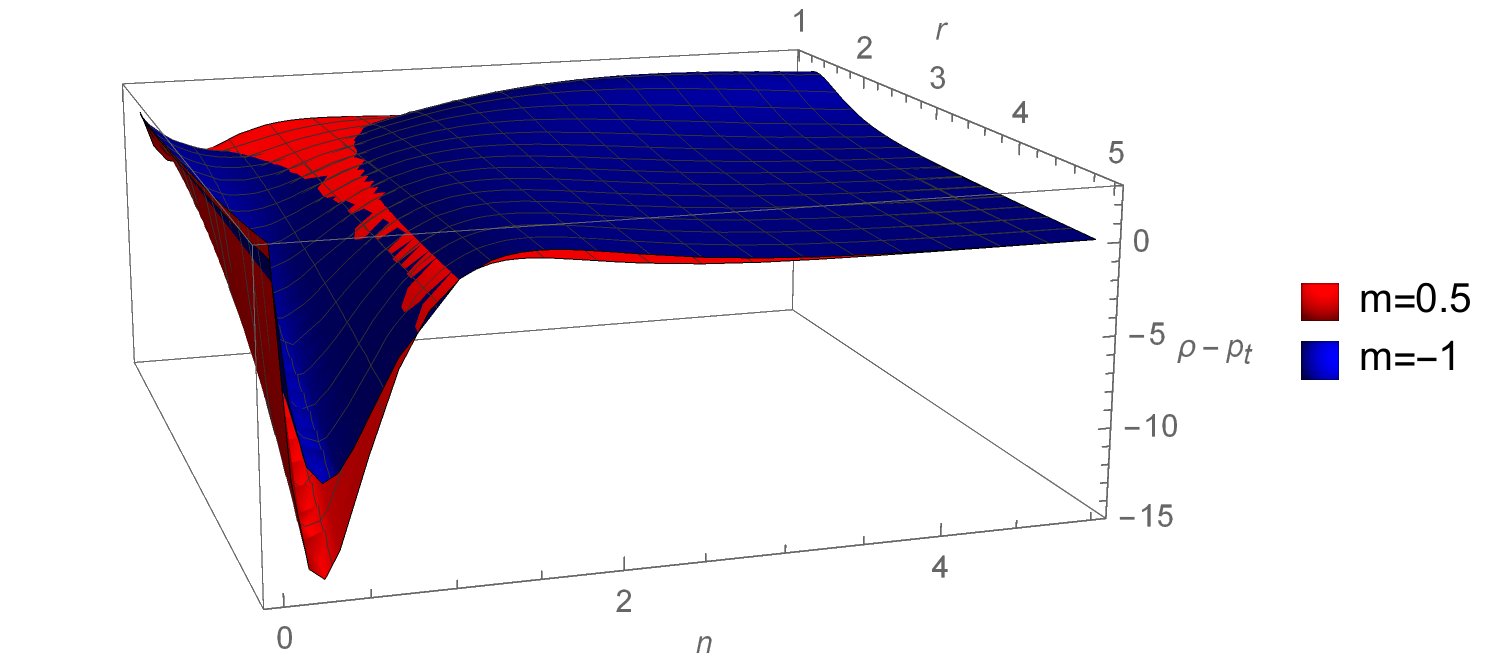}}
    \subfloat{\includegraphics[width=0.5\hsize]{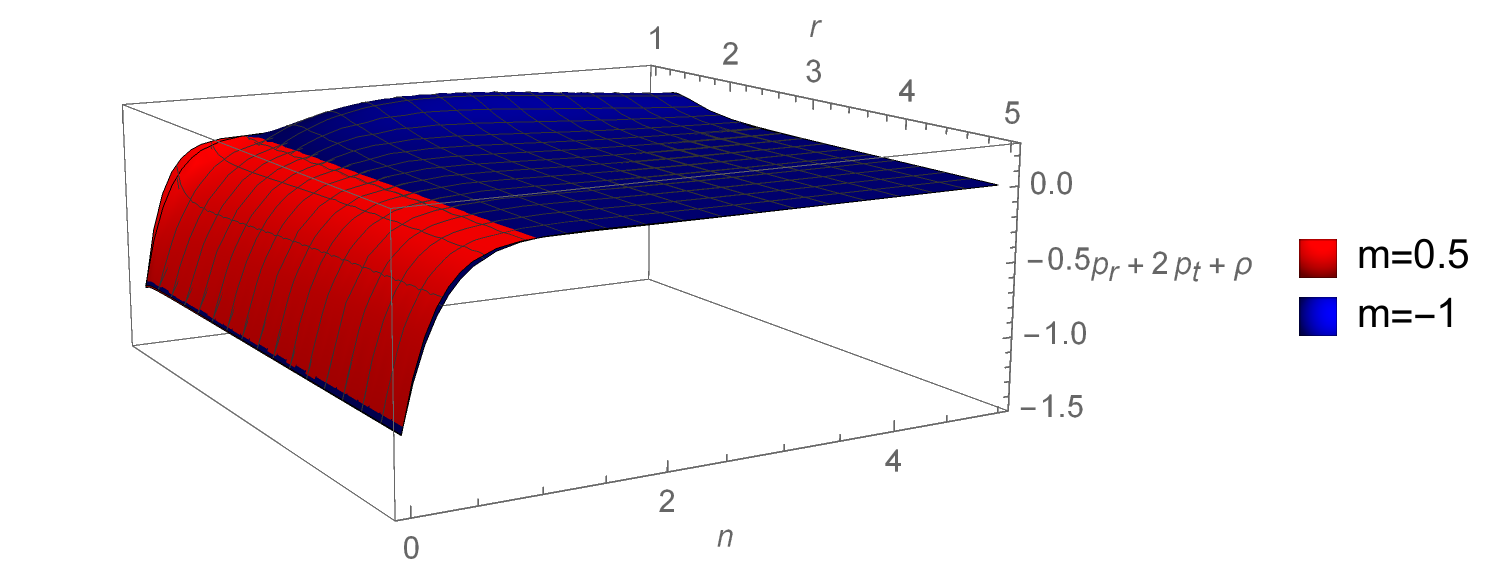}}
    \caption{Three-dimensional plot of the energy conditions as functions of the radial coordinate $r$ and the parameter $n$ for a power-law $f(Q)$ model with the shape function  $b(r)=r_0\left(\frac{r}{r_0}\right)^{m}$.} \label{fig:4.2}
\end{figure}

 \Cref{fig:4.1} reveals that by selecting appropriate values for the parameter $n$ (e.g. $n=3, 4$), the matter in traversable wormhole can satisfy all four energy conditions, regardless of whether $m$ is positive ($m=1/2$) or negative ($m=-1$). We can also observe that when the parameters are set to $m=-1$ and $n=2$, the weak energy condition of matter does not hold in the vicinity of the wormhole throat. To further investigate the impact of parameter values $n$ on the validity of energy conditions for wormhole matter under the $f(Q)$ model, we plot in \Cref{fig:4.2} the three-dimensional pictures showing the variation of ECs  with respect to the radial coordinate $r$ and parameter $n$.

 From \Cref{fig:4.2} (the three-dimensional pictures), it can be observed that under the conditions of shape functions $b(r)=r_0\left(\frac{r}{r_0}\right)^{m}$ and $a=1$, the same conclusion can be reached regardless of parameters $m=1/2$ or $m=-1$. When $n>1$, traversable wormhole matter satisfies all four energy conditions; when $0 \leq n \leq 1$, traversable wormhole matter does not satisfy any energy condition. Since parameter $a$ plays the role of a proportionality parameter in equations (\ref{4.12})-(\ref{4.17}), it is evident that when $a=-1$, the conclusions obtained are completely opposite to the case when $a=1$. Therefore, our researches reveal that in the $f(Q)$ model and the given shape function, the satisfaction of the four energy conditions for traversable wormhole matter is not sensitive to the values of model parameter $n$. Within a broad range of parameter values of $n$, the energy conditions for wormhole matter maintain valid.

\subsection*{\text{Shape function: $b(r)=\frac{r_0 \ln (r+1)}{\ln \left(r_0+1\right)}$}}

\begin{figure}[ht]
    \renewcommand{\thefigure}{3}
    \centering
    \subfloat{\includegraphics[width=0.5\hsize]{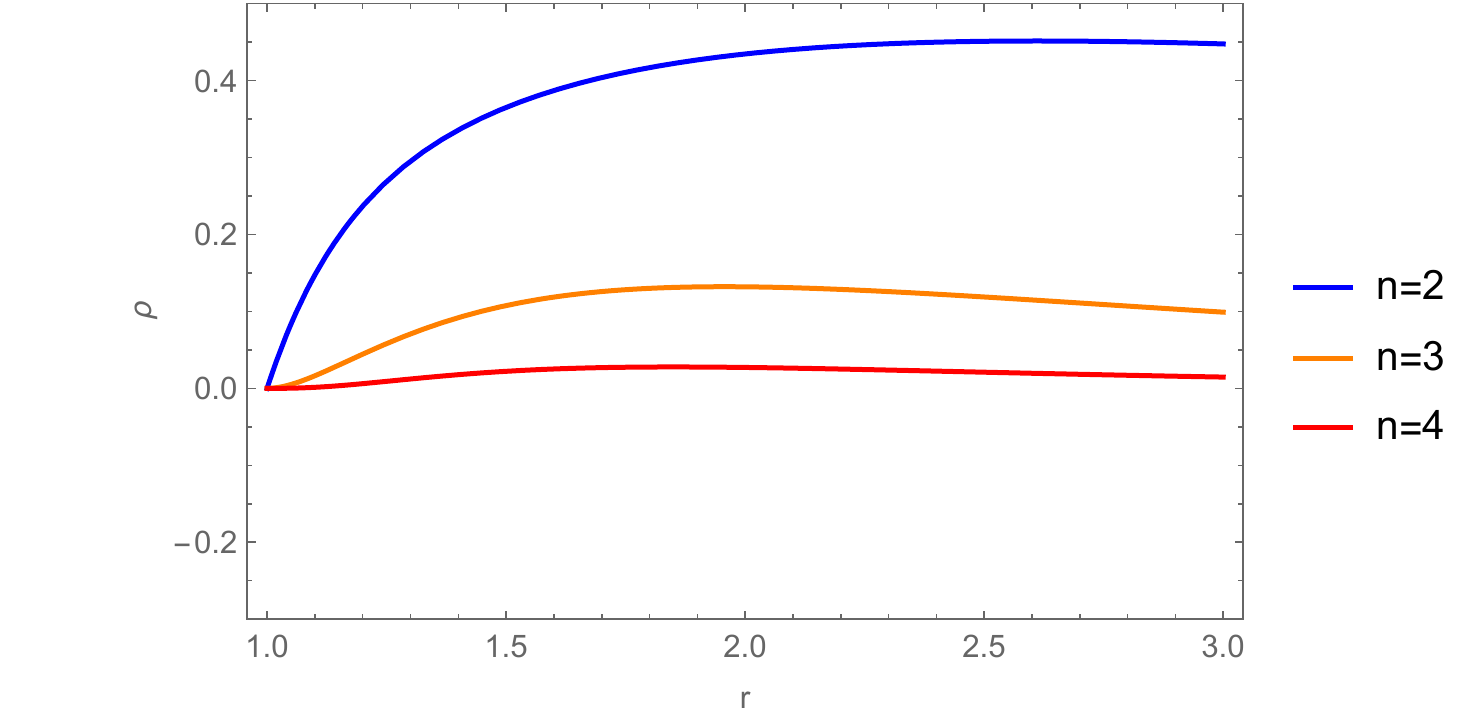}}
    \subfloat{\includegraphics[width=0.5\hsize]{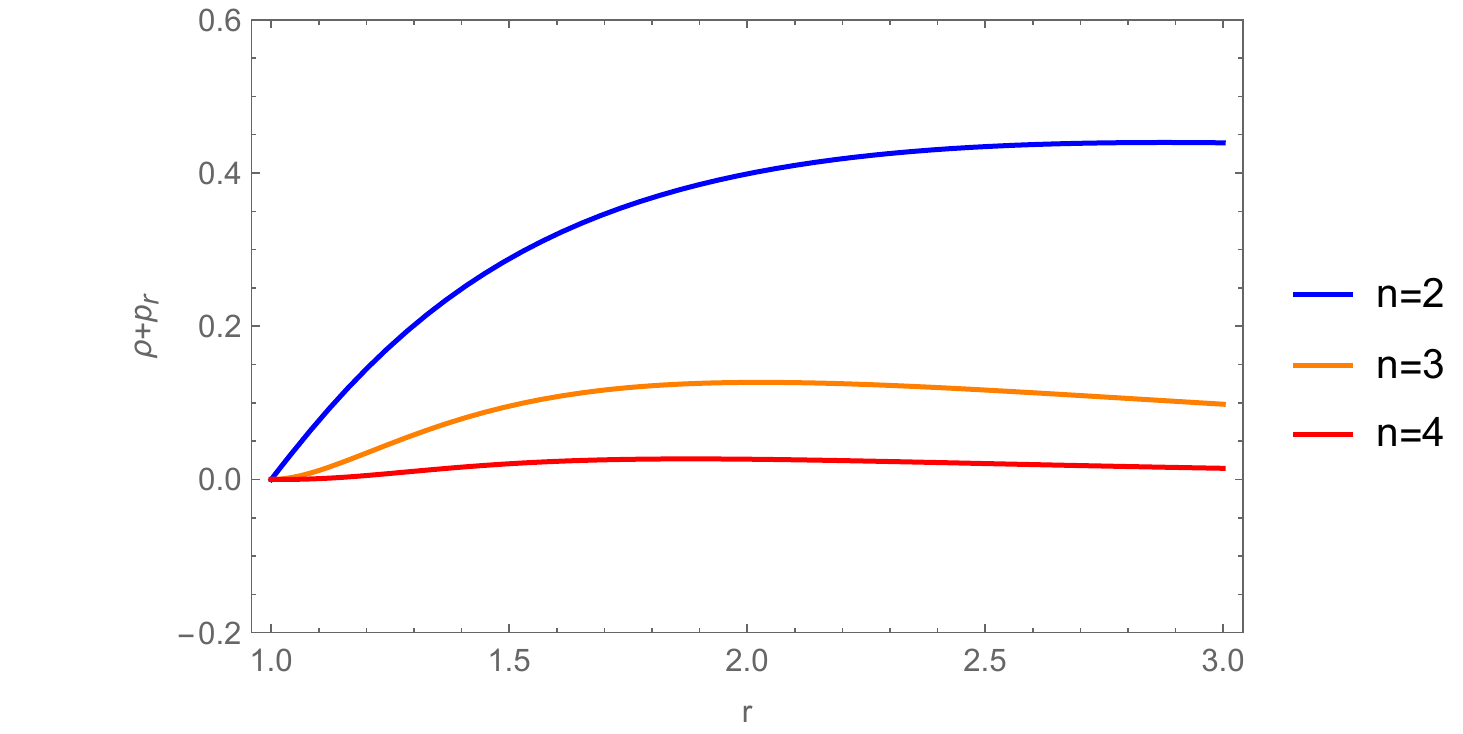}}
    \\
    \subfloat{\includegraphics[width=0.5\hsize]{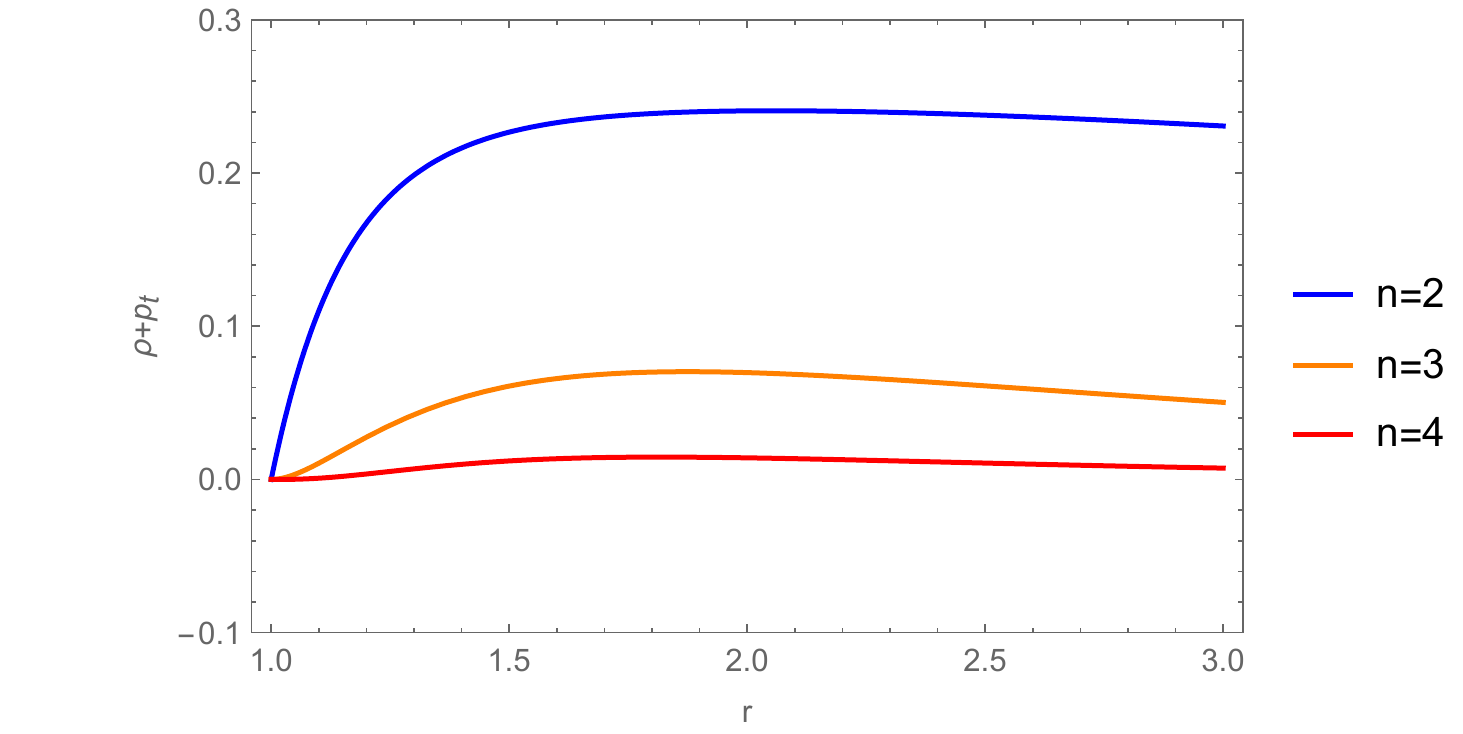}} 
    \subfloat{\includegraphics[width=0.5\hsize]{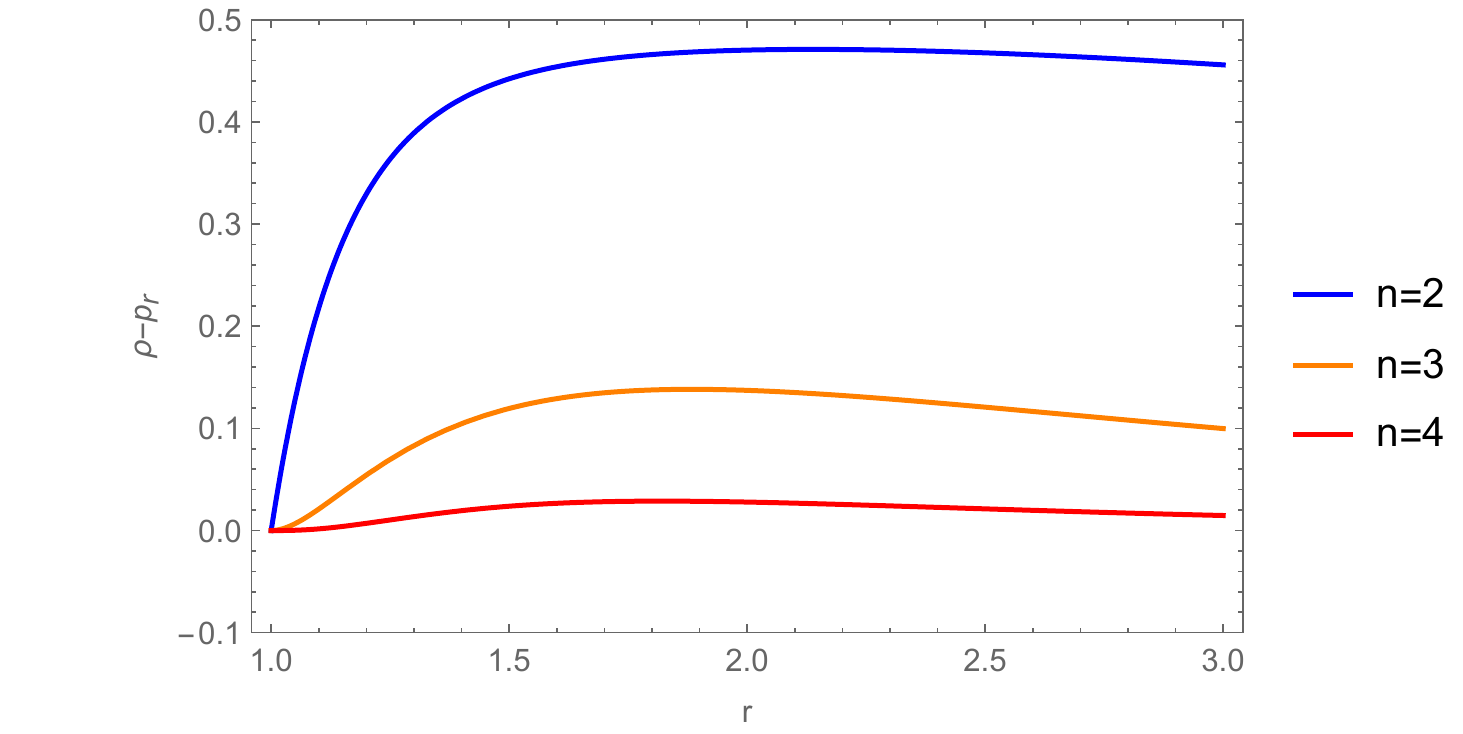}}
    \\
    \subfloat{\includegraphics[width=0.5\hsize]{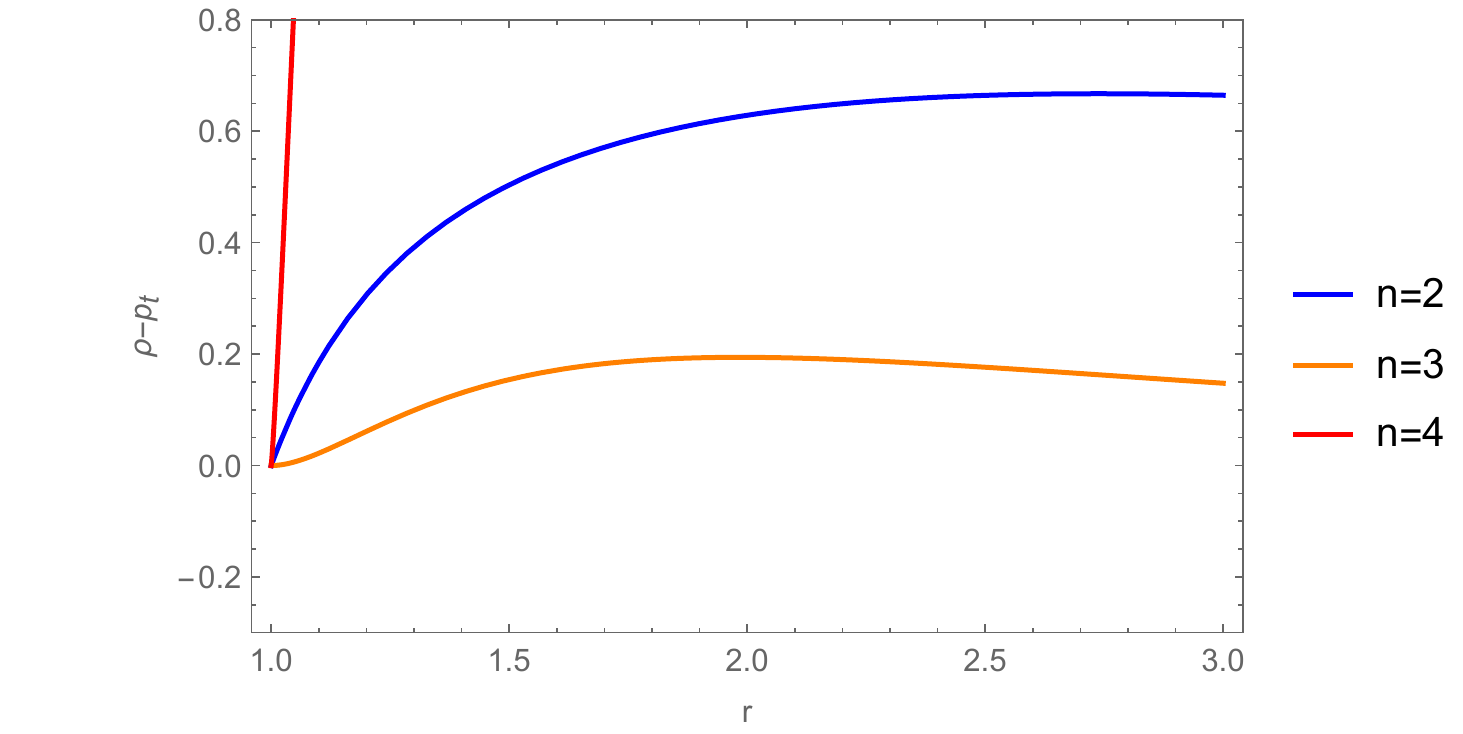}}
    \subfloat{\includegraphics[width=0.5\hsize]{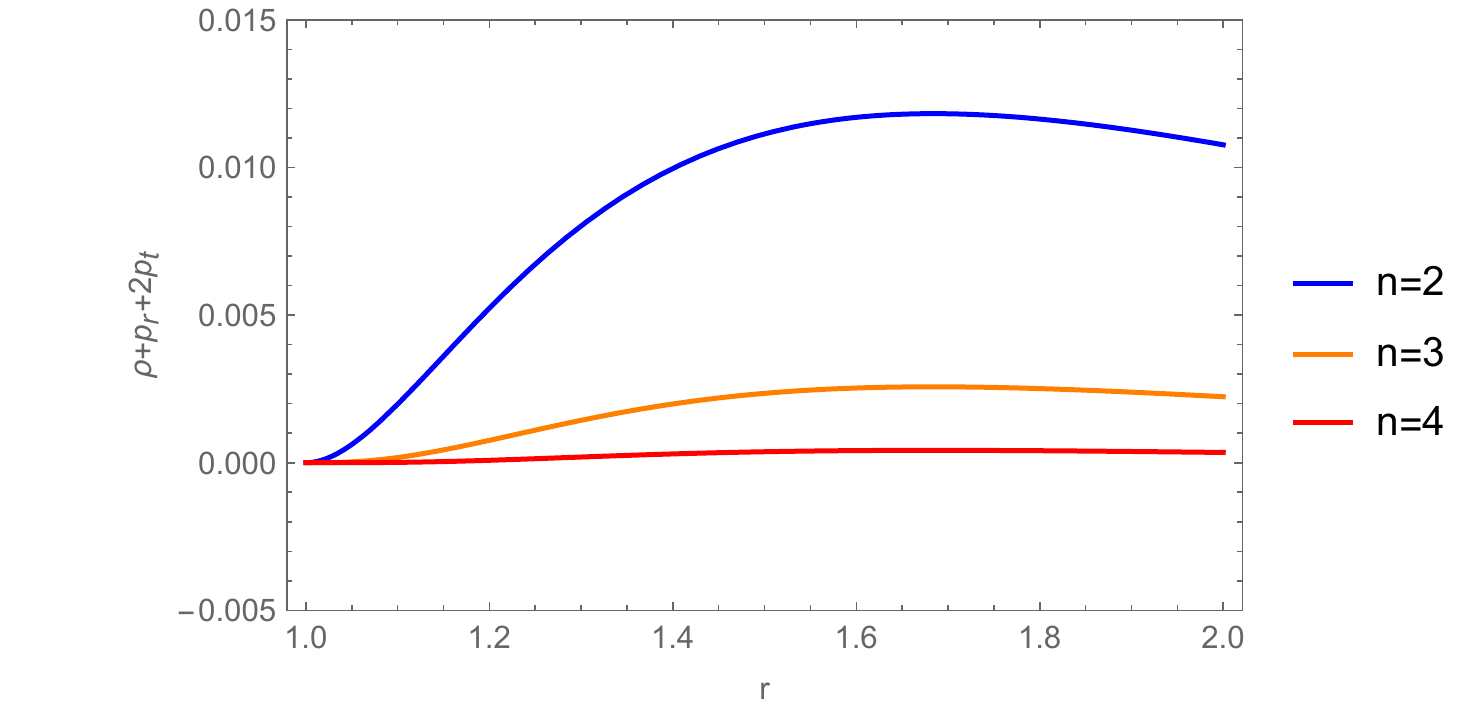}}
    \caption{Variation of the energy conditions with respect to the radial coordinate $r$ in the $f(Q)$ power-law function model, under the shape function  $b(r)=\frac{r_0 \ln (r+1)}{\ln \left(r_0+1\right)}$.} \label{fig:4.3}
\end{figure}

 Next, within the framework of symmetric parallel gravitational model $f(Q)=a(-Q)^n$, we discuss the influence of different shape function on the physical properties of wormholes. Here, we consider selecting the shape function: $b(r)=\frac{r_0 \ln (r+1)}{\ln \left(r_0+1\right)}$ \cite{julianto2022traversable,sharma2022traversable,godani2019traversable}. Compared to the above shape function, this model of shape function contains only one model parameter $r_0$, which makes the model more simple. Substituting the shape function $b(r)=\frac{r_0 \ln (r+1)}{\ln \left(r_0+1\right)}$  into equations (\ref{4.9})-(\ref{4.11}), we receive the specific expressions for the four energy conditions under this case as follows:
\begin{equation}
\begin{gathered}
\rho=  2^{n-1} a\left(\frac{r-\frac{r_0 \ln (1+r)}{\ln \left(1+r_0\right)}}{r^3}\right)^{n-1}\left[n r(n-1)+\frac{r-n r+\frac{r_0[\mathrm{nr}+(n-1)(1+r) \ln (1+r)]}{(1+r) \ln \left(1+r_0\right)}}{r^3}\right],\label{4.18}
\end{gathered}
\end{equation}

\begin{equation}
\small
\begin{gathered}
\rho+P_r=      \frac{2^{n-1} a\left\{(n-1) n r^4(1+r) \ln \left(1+r_0\right)+[n r-n(1+r) \ln (1+r)] r_0\right\}\left(\frac{r-\frac{\ln (1+r) r_0}{\ln \left(1+r_0\right)}}{r^3}\right)^n}{(1+r)\left(r \ln \left(1+r_0\right)-\ln (1+r) r_0\right)},\label{4.19}
\end{gathered}
\end{equation}

\begin{equation}
\small
\begin{gathered}
\rho+P_t=   \frac{2^{n-2} a\left\{(n-1) n r^4(1+r) \ln \left(1+r_0\right)+[n r+n(1+r) \ln (1+r)] r_0\right\}\left(\frac{r-\frac{\ln (1+r) r_0}{\ln \left(1+r_0\right)}}{r^3}\right)^n}{(1+r)\left(r \ln \left(1+r_0\right)-\ln (1+r) r_0\right)},\label{4.20}
\end{gathered}
\end{equation}

\begin{equation}
\small
\begin{gathered}
\rho-P_r=    \frac{2^{n-1} a\left\{(n-1) r(1+r)\left(n r^3-2\right) \ln \left(1+r_0\right)+[\mathrm{nr}+(3 n-2)(1+r) \ln (1+r)] r_0\right\}\left(\frac{r-\frac{\ln (1+r) r_0}{\ln \left(1+r_0\right)}}{r^3}\right)^n}{(1+r)\left(r \ln \left(1+r_0\right)-\ln (1+r) r_0\right)},\label{4.21}
\end{gathered}
\end{equation}
   
\begin{equation}
\small
\begin{gathered}
\rho-P_t=   \frac{2^{n-2} a\left\{(n-1) r(1+r)\left(3 n r^3-4\right) \ln \left(1+r_0\right)+[3 n r+(3 n-4)(1+r) \ln (1+r)] r_0\right\}\left(\frac{r-\frac{\ln (1+r) r_0}{\ln \left(1+r_0\right)}}{r^3}\right)^n}{(1+r)\left(r \ln \left(1+r_0\right)-\ln (1+r) r_0\right)},\label{4.22}
\end{gathered}
\end{equation}
 
\begin{equation}
\rho+P_r+2 P_t=2^n a(n-1)\left(\frac{r-\frac{r_0 \ln (1+r)}{\ln \left(1+r_0\right)}}{r^3}\right)^n.\label{4.23}
\end{equation}
 Using equations (\ref{4.18})-(\ref{4.23}), we can numerically analyze the energy conditions and related properties of the traversable wormhole material under this condition. The results of numerical calculations are shown in \Cref{fig:4.3}. Similar to the study in Section I, it is evident that parameter $a$ in equations (\ref{4.18})-(\ref{4.23}) merely serves as a proportionality parameter. Its value only affects the magnitude of the energy condition functions, without altering the positivity or negativity of the energy conditions. Therefore, without loss of generality, we choose the constant parameter $a=1$.

\begin{figure}[ht]
    \renewcommand{\thefigure}{4}
    \centering
    \subfloat{\includegraphics[width=0.5\hsize]{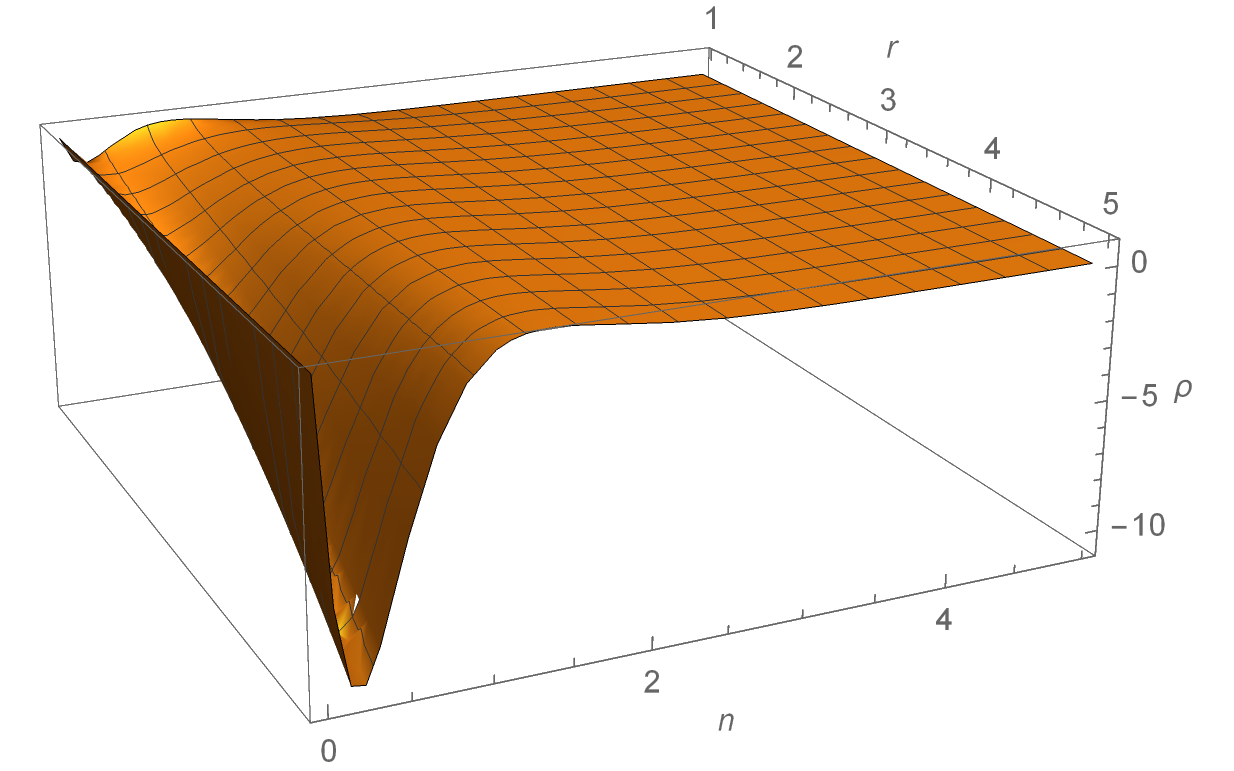}}
    \subfloat{\includegraphics[width=0.5\hsize]{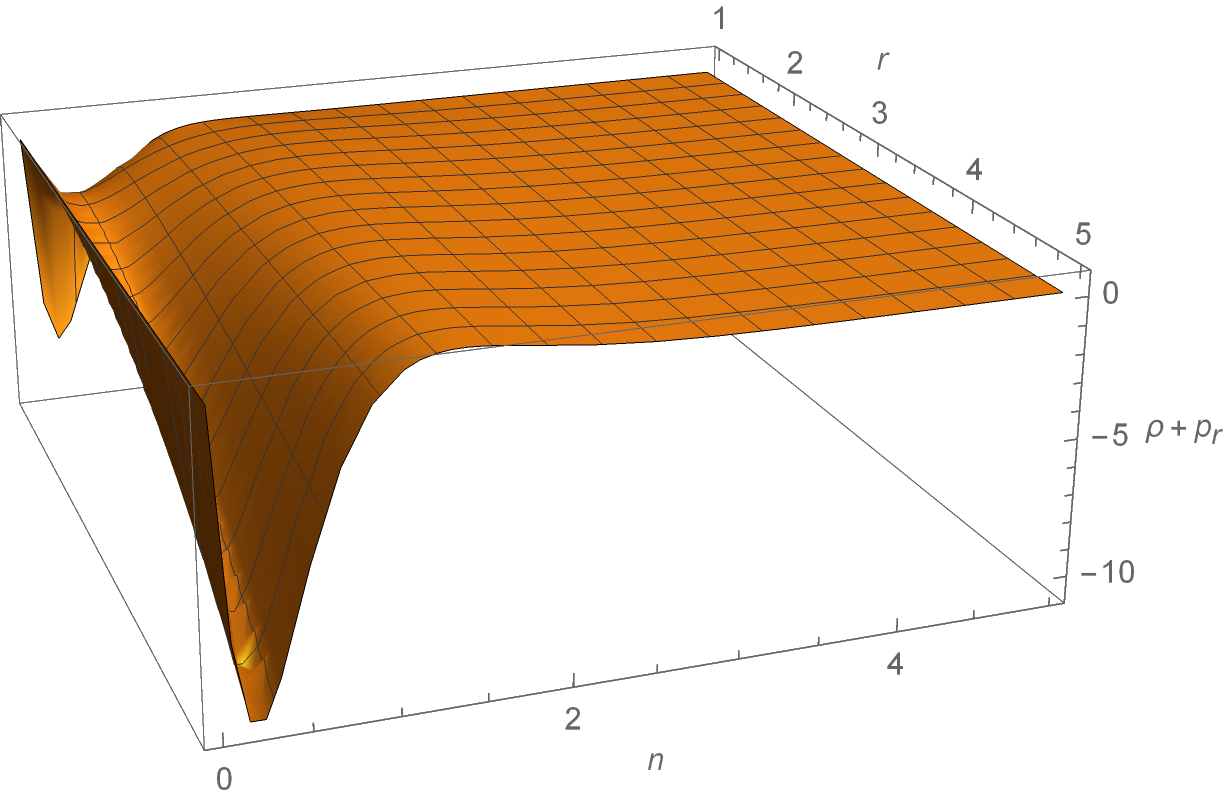}}\\
    \subfloat{\includegraphics[width=0.5\hsize]{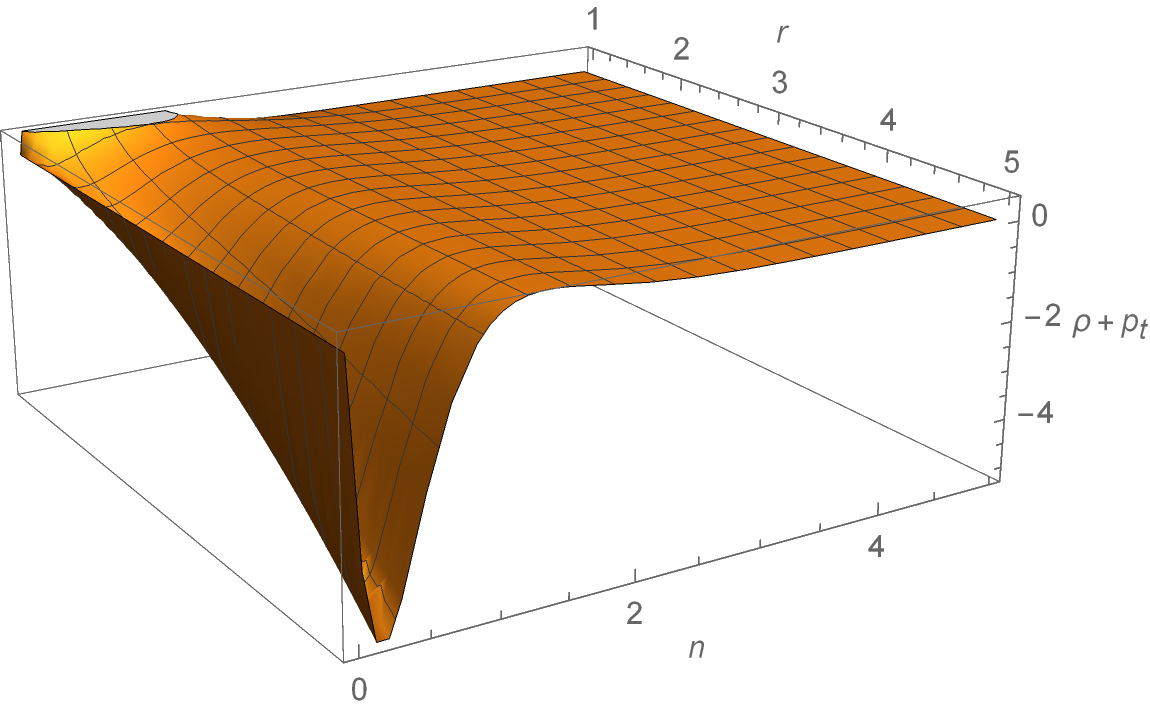}} 
    \subfloat{\includegraphics[width=0.5\hsize]{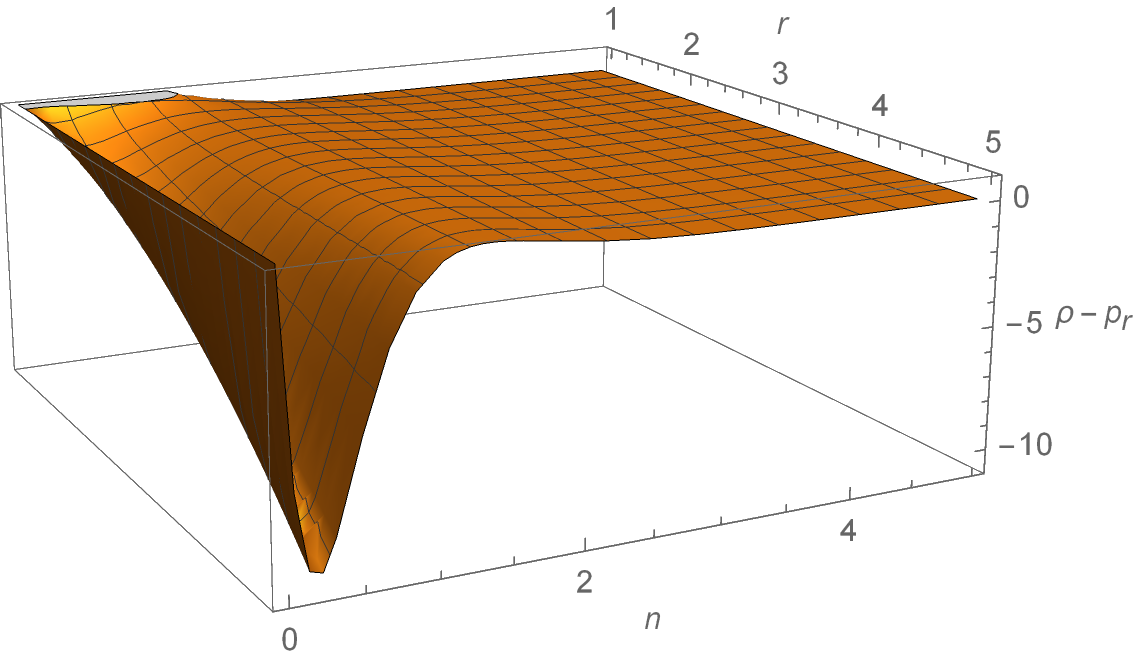}}\\
    \subfloat{\includegraphics[width=0.5\hsize]{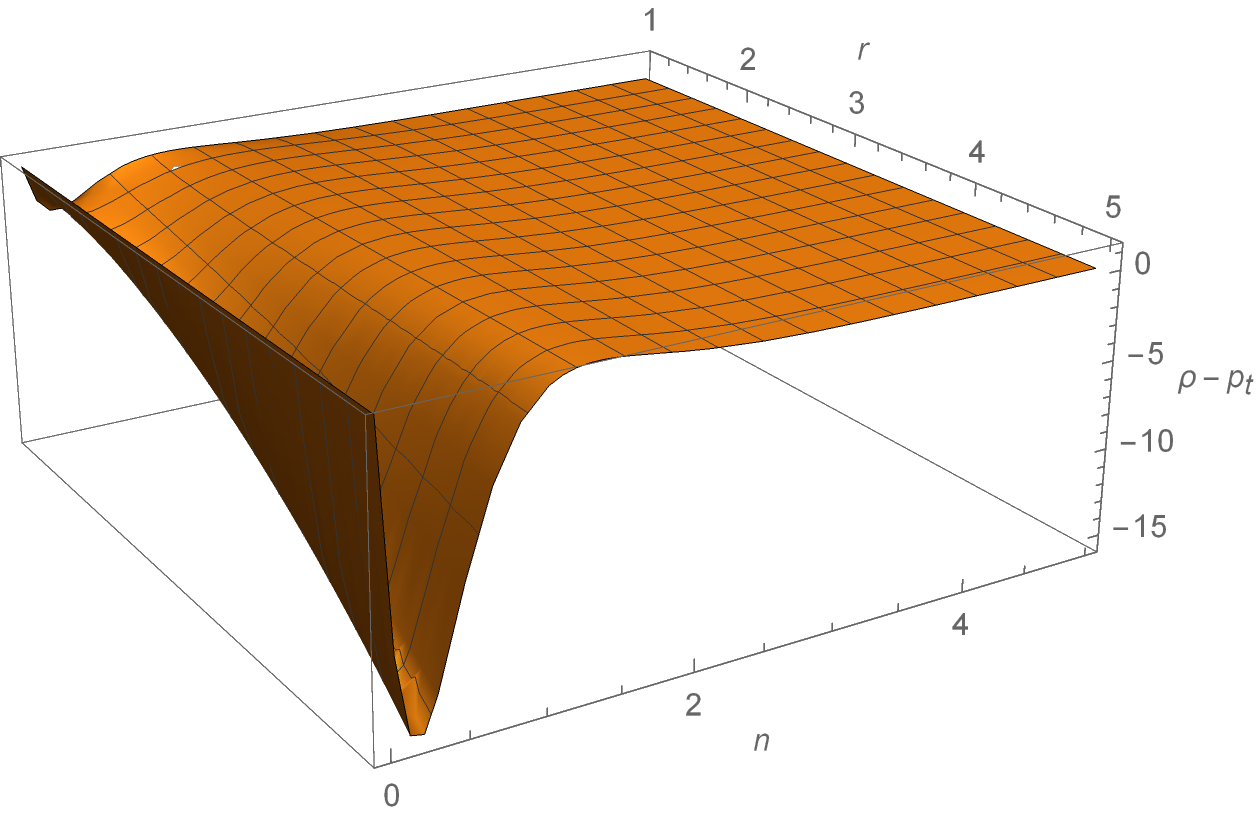}}
    \subfloat{\includegraphics[width=0.5\hsize]{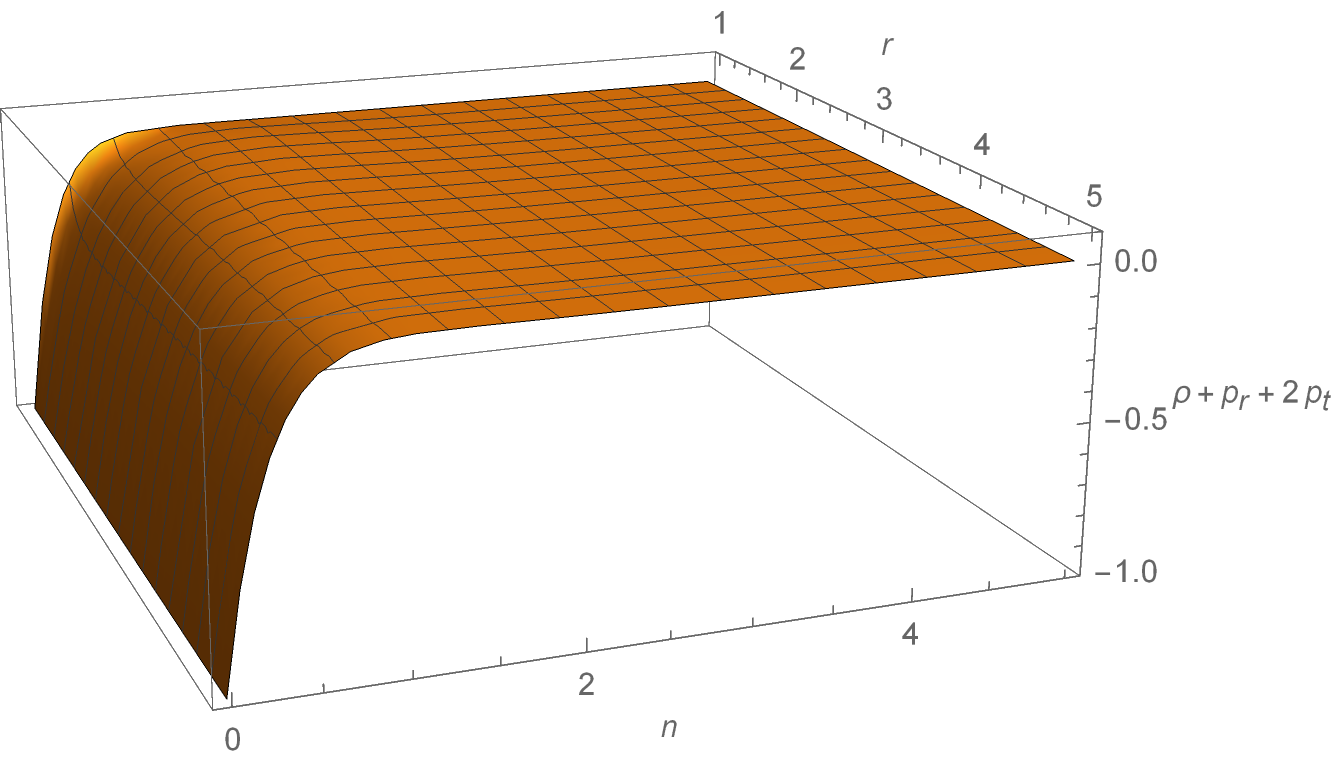}}
    \caption{ In the $f(Q)$ power-law model, with the shape function  $b(r)=\frac{r_0 \ln (r+1)}{\ln \left(r_0+1\right)}$, the three-dimensional pictures of energy conditions in relation to the radial coordinate $r$ and the variation of the parameter $n$.} \label{fig:4.4}
 \end{figure}

 From \Cref{fig:4.3}, we can see that for the discussed $f(Q)$ power-law model, when the shape function is $b(r)=\frac{r_0 \ln (r+1)}{\ln \left(r_0+1\right)}$, the four energy conditions for the matter inside the wormhole can all be satisfied. Obviously, in \Cref{fig:4.3}, we only consider the validity of the energy conditions when the parameter $n$ takes some discrete values ($n=2,3,4$). To explore the impact of the parameter $n$ in more general cases on the energy conditions, we plot the three-dimensional image of the energy conditions with respect to the radial coordinate $r$ and the parameter $n$ in \Cref{fig:4.4}.

 From \Cref{fig:4.4}, the following conclusions can be drawn under the conditions of shape function $b(r)=\frac{r_0 \ln (r+1)}{\ln \left(r_0+1\right)}$ and $a=1$. If $n>1$, all four energy conditions for traversable wormhole matter are satisfied. If $0 \leq n \leq 1$, none of the four energy conditions are satisfied. However, for the case where parameter $a=-1$, equations (\ref{4.18})-(\ref{4.23}) yield conclusions opposite to the case where $a=1$. Specifically, when $n>1$, none of the four energy conditions for traversable wormhole matter are satisfied, whereas when $0 \leq n \leq 1$, all four energy conditions are satisfied.

\begin{figure}[ht]
    \renewcommand{\thefigure}{5}
    \centering
    \subfloat{\includegraphics[width=0.5\hsize]{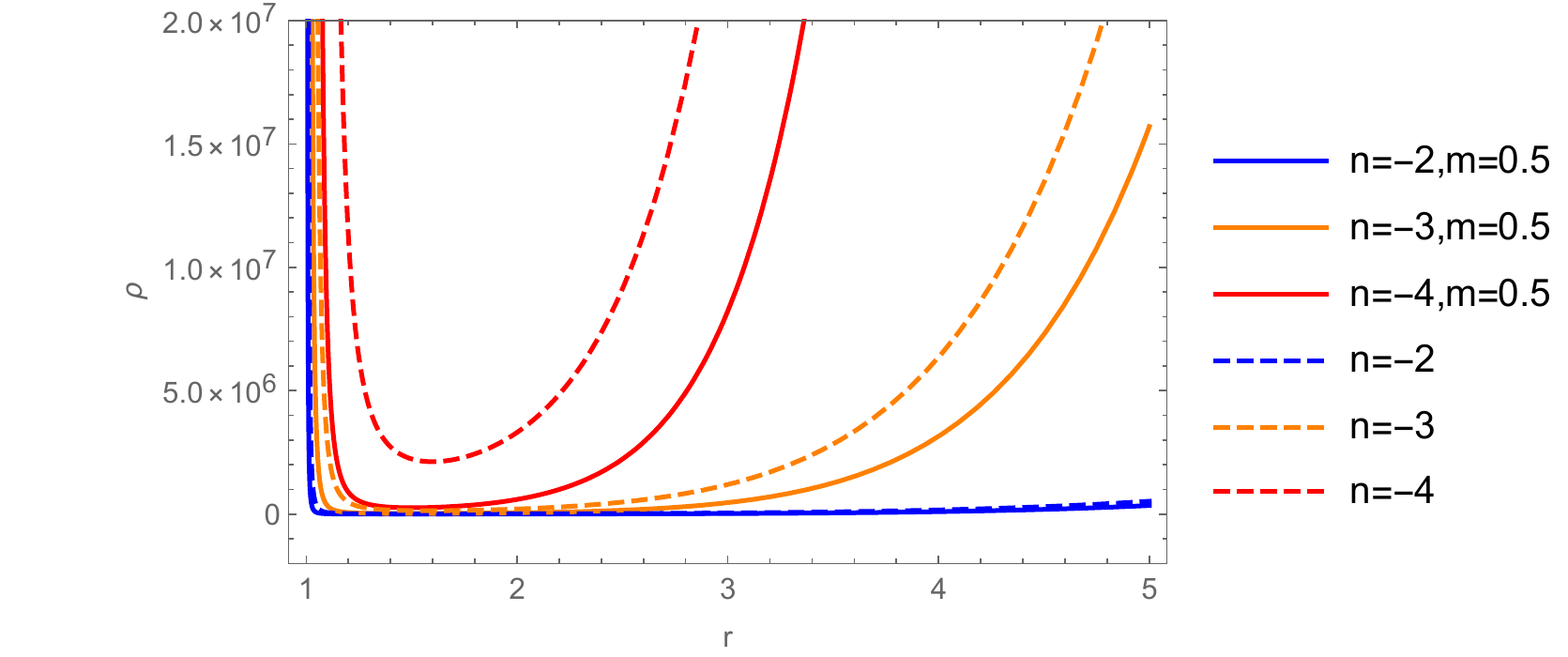}}
    \subfloat{\includegraphics[width=0.5\hsize]{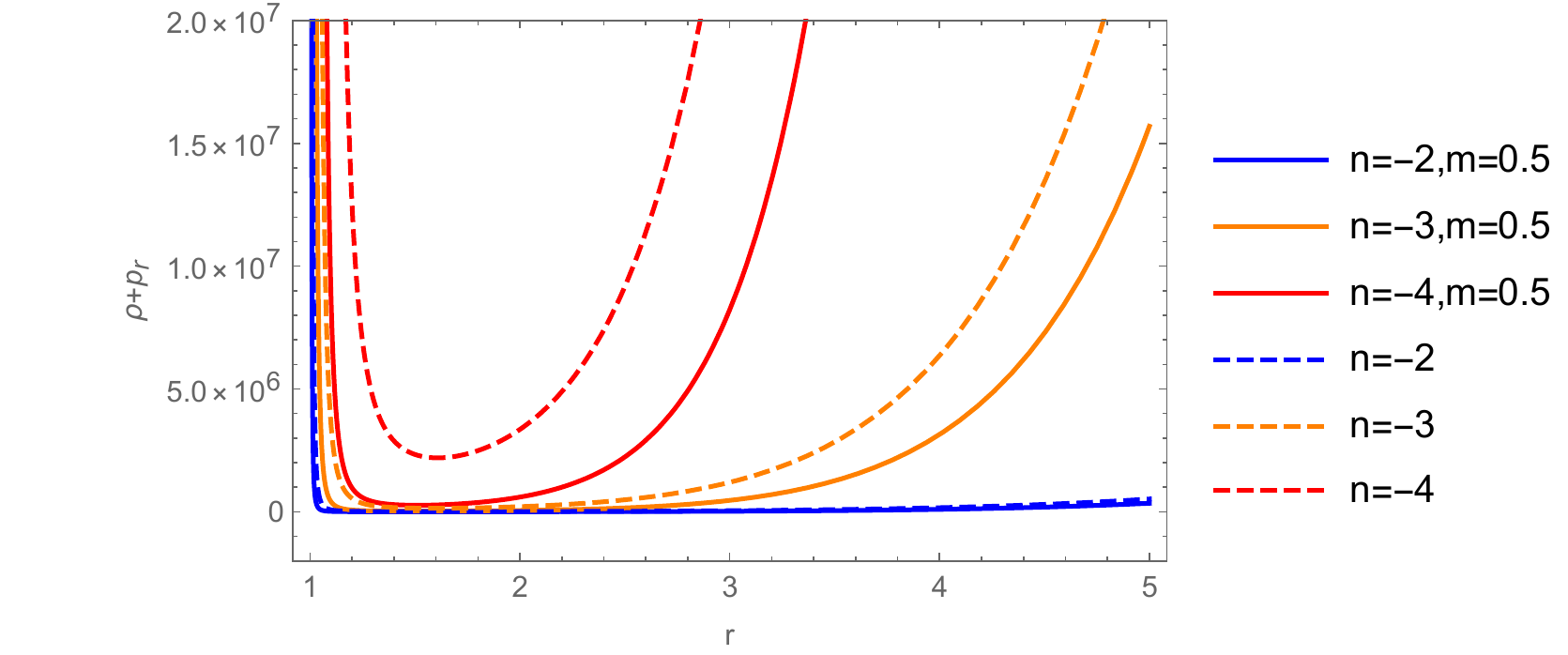}}\\
    \subfloat{\includegraphics[width=0.5\hsize]{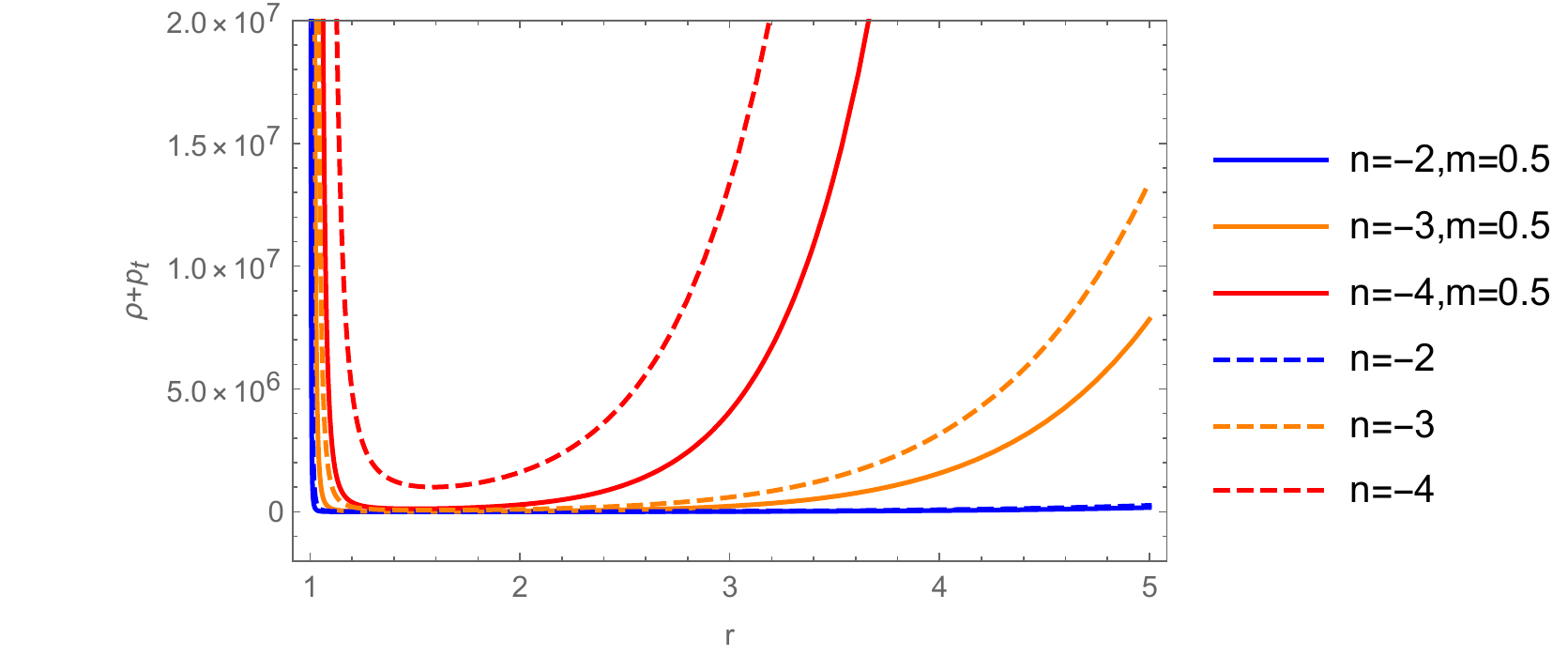}} 
    \subfloat{\includegraphics[width=0.5\hsize]{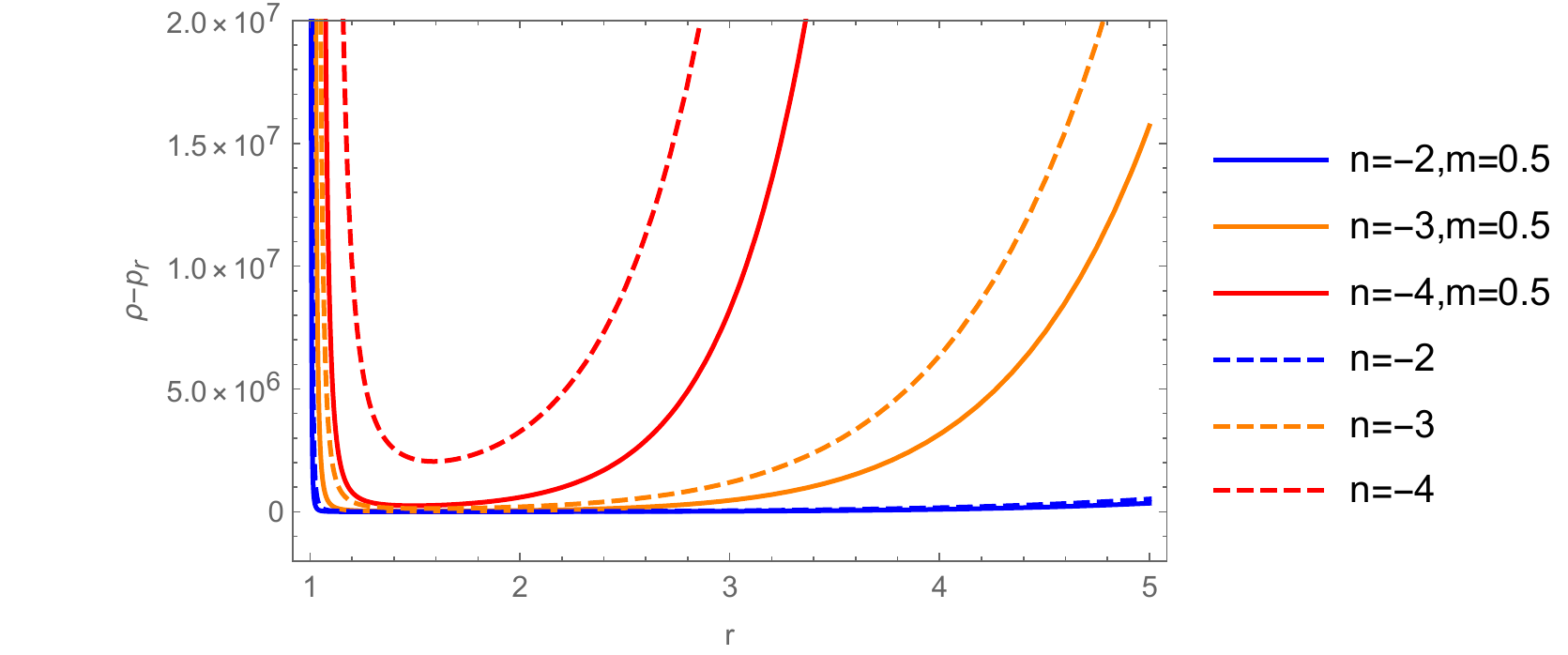}}\\
    \subfloat{\includegraphics[width=0.5\hsize]{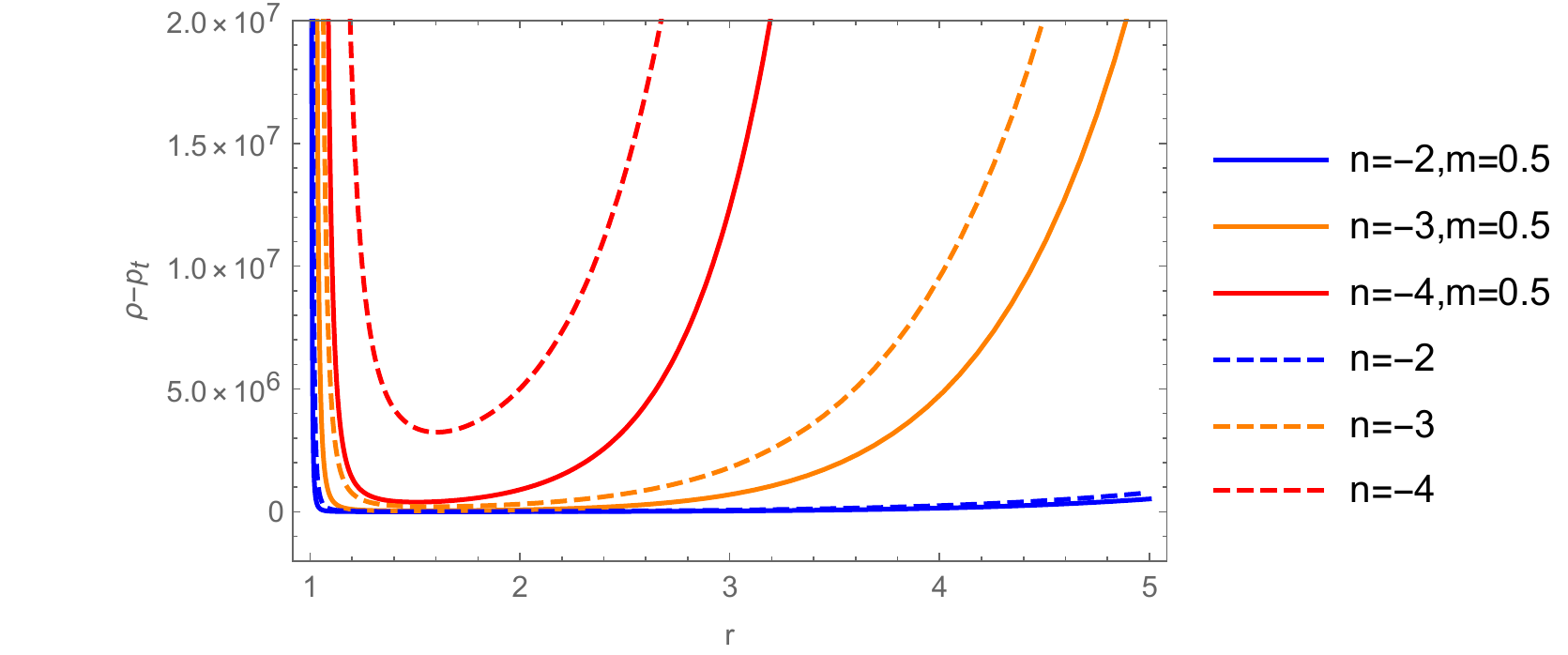}}
    \subfloat{\includegraphics[width=0.5\hsize]{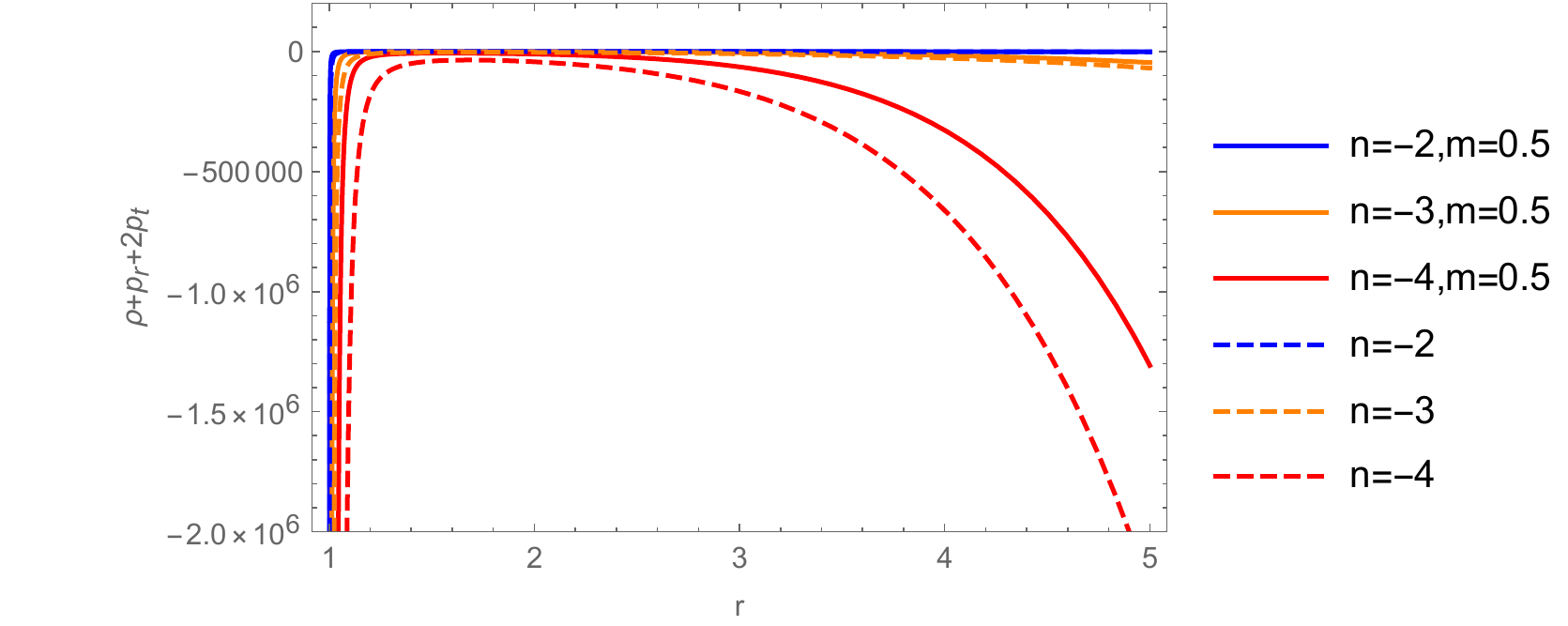}}
    \caption{ In the $f(Q)$ power-law function model, the change in the energy conditions relative to the radial coordinate $r$ is depicted. Solid lines represent the plot for the shape function $b(r)=r_0\left(\frac{r}{r_0}\right)^m$ ($m=0.5$), while dashed lines correspond to the plot for the shape function $b(r)=\frac{r_0 \ln (r+1)}{\ln \left(r_0+1\right)}$.}\label{fig:4.5}
\end{figure}

 For comparison, we have also plotted the evolution of the material energy conditions within the wormhole relative to the radial coordinate $r$, when $n<0$, for the two shape functions $b(r)=r_0\left(\frac{r}{r_0}\right)^m$ and $b(r)=\frac{r_0 \ln (r+1)}{\ln \left(r_0+1\right)}$ (as shown in \Cref{fig:4.5}). \Cref{fig:4.5} indicates that in the $f(Q)=a(-Q)^n$ model, when $a=1$ and $n<0$, the matter of traversable wormholes can satisfy the NEC, the WEC, and the DEC under both shape functions. However, it does not satisfy the SEC.

 From \Cref{fig:4.1,fig:4.2,fig:4.3,fig:4.4,fig:4.5}, we observe (taking the case of constant parameter $a=1$ as an example) that for the power-law model of $f(Q)$ ($f(Q)=a(-Q)^n$), under the two types of shape functions considered, the following conclusions can be drawn: If $n>1$, traversable wormhole matter satisfies all four energy conditions. If $0 \leq n \leq 1$, traversable wormhole matter does not satisfy any of the energy conditions. If $n<0$, traversable wormhole matter satisfies the null energy condition, weak  energy condition, and dominant energy condition, but violates the strong  energy condition.

\section*{Conclusion}\label{sec5}

In this paper, we investigated the properties of static, spherically symmetric traversable wormholes within the framework of $f(Q)$ modified gravity theory. By considering two different shape functions, we discussed the validity of energy conditions for matter within the wormhole geometry under the $f(Q)$ power-law model. Our studies revealed that traversable wormholes can be realized without the need for exotic or special matter under the power-law model: $f(Q)=a(-Q)^n$, as conventional matter satisfies all four energy conditions. Specifically, within the chosen power-law model of $f(Q)$, we considered two different shape functions. When $a=1$ and $n>1$, or $a=-1$ and $0 \leq n \leq 1$, all four energy conditions for traversable wormhole matter are satisfied. This implies that we can obtain stable traversable wormholes without the introduction of any exotic or special matter. Conversely, when $a=1$ and $0 \leq n \leq 1$, or $a=-1$ and $n>1$, none of the four energy conditions for traversable wormhole matter are satisfied. For the case of $a=1$ and $n<0$, the energy conditions of matter in the wormhole are not fully satisfied (satisfying NEC, WEC, DEC but not SEC). In conclusion, under the assumption of model parameters $a=1$ and $n>1$, or $a=-1$ and $0 \leq n \leq 1$, there exist gravitational theory model within the framework of modified gravitational theory  that allow stable traversable wormholes without the need for exotic matter, e.g., the $f(Q)$ power-law theoretical model.

\bmhead{Acknowledgements}

The research work is supported by the National Natural Science Foundation of China (12175095,12075109 and 11865012), and supported by  LiaoNing Revitalization Talents Program (XLYC2007047).

\bmhead{Data Availability Statement}   
No Data associated in the manuscript.


\bibliography{sn-bibliography}

\end{document}